\documentclass[twocolumn]{aastex631}

\defcitealias{2003MNRAS.344.1000B}{BC03}

\begin{document}

\title{Spectroscopic study of globular and fuzzy clusters in Lenticular galaxy NGC\,1023}

\author[0009-0005-0469-3590]{M. A. López-Santamaría}
\affiliation{Instituto Nacional de Astrofísica, Óptica y Electrónica Luis Enrique Erro 1, Tonantzintla, 72840, Puebla, Mexico}

\author[0000-0002-4677-0516]{Y. D. Mayya}
\affiliation{Instituto Nacional de Astrofísica, Óptica y Electrónica Luis Enrique Erro 1, Tonantzintla, 72840, Puebla, Mexico}

\author[0000-0003-2127-2841]{Luis Lomelí-Núñez}
\affiliation{Valongo Observatory, Federal University of Rio de Janeiro, 
Ladeira Pedro Antonio 43, Saude Rio de Janeiro, RJ 20080-090, Brazil}

\author{L. H. Rodríguez-Merino}
\affiliation{Instituto Nacional de Astrofísica, Óptica y Electrónica Luis Enrique Erro 1, Tonantzintla, 72840, Puebla, Mexico}

\author[0000-0002-8351-8854]{Jairo A. Alzate}
\affiliation{Centro de Estudios de Física del Cosmos de Aragón, Teruel, 44001, Spain}

\author{Arianna Cortesi}
\affiliation{Valongo Observatory, Federal University of Rio de Janeiro, 
Ladeira Pedro Antonio 43, Saude Rio de Janeiro, RJ 20080-090, Brazil}

\author{P. A. Ovando}\affiliation{Instituto Nacional de Astrofísica, Óptica y Electrónica Luis Enrique Erro 1, Tonantzintla, 72840, Puebla, Mexico}


\author[0000-0003-1327-0838]{D. Rosa-González}
\affiliation{Instituto Nacional de Astrofísica, Óptica y Electrónica Luis Enrique Erro 1, Tonantzintla, 72840, Puebla, Mexico}



\begin{abstract}

We here report the results from spectroscopic observations of a sample of 26 globular cluster (GC) and 21 faint fuzzy (FF) candidates in the lenticular galaxy NGC\,1023 using the 10.4-m Gran Telescopio Canarias. Using the recessional velocities and stellar absorption features, we determine that 18 and 9 of the observed candidates are bona fide GCs and FFs, respectively. The majority of the rejected FF candidates are background emission line galaxies for which we determine their redshifts.
We used the spectroscopic data to determine velocity, age, metallicity and extinction of all bona fide clusters. We find that FFs are clearly younger (age~=~7--9~Gyr) than GCs (age $>$ 10~Gyr). Both kind of clusters in this galaxy are metal-rich ([Fe/H] = $-0.58~\pm$ 0.33). The  ages and metallicities of individual FFs reported here are the first such measurements in any galaxy and agree with the previously-reported measurement on stacked spectrum. The kinematical analysis reaffirms that the FFs belong to the disk of the galaxy, suggesting that their progenitors are most likely massive, compact disk clusters that have been able to survive for long timescales. We propose that the fuzzy appearance of FFs as compared to the GCs is a consequence of the dynamical evolution of their progenitor super star clusters in the disks of low-mass galaxies.

\end{abstract}



\section{Introduction} \label{sec:intro}

\begin{figure*}
\label{fig:HST}\includegraphics[width=\textwidth]{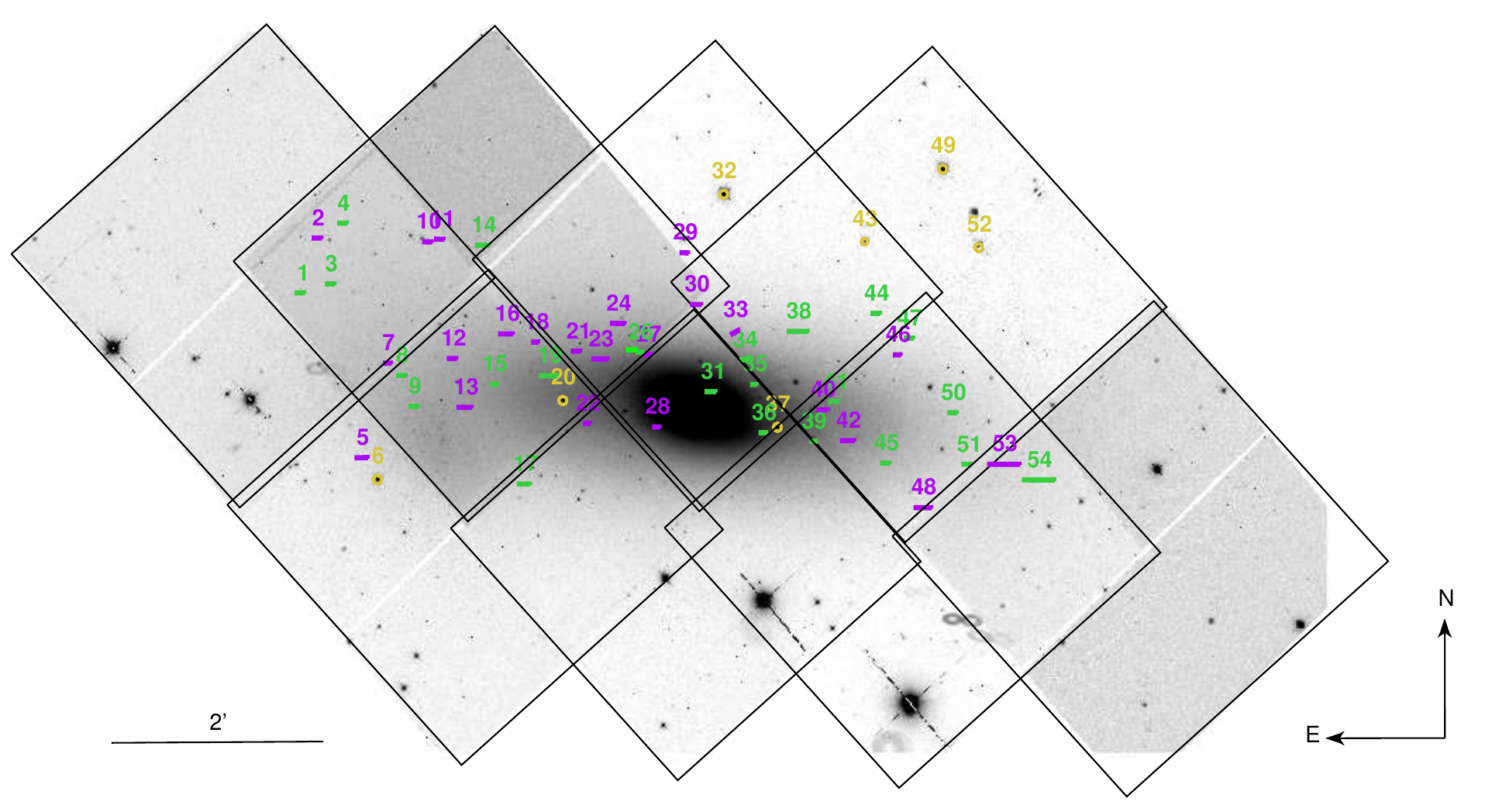}
\caption{Mosaic of the Lenticular galaxy NGC 1023 (RA = 02h 40m 24.008s, Dec = +39d 03m 47.888s) in the \textit{F}814\textit{W} filter, constructed from eight individual fields of the \textit{HST}/ACS. This image displays the positions of the 54 slitlets, numbered sequentially from east to west, placed on the globular (green rectangles), faint fuzzy clusters (violet rectangles) and fiducial stars (yellow circles) used in all our spectroscopic observations.}
\label{fig:HST}
\end{figure*}

Globular Clusters (GCs) are old systems, some of the earliest-formed objects in the Universe, and are known for their high density of stars and stability through time. GCs are seen as fossil records since they preserve the characteristics of the medium in which they formed, holding crucial information regarding the formation and evolution of galaxies \citep{2020rfma.book..245B}.

\textit{Hubble Space Telescope} (\textit{HST}) has allowed the discovery of a new class of clusters known as faint fuzzy clusters or simply faint fuzzies (FF). These clusters were detected for the first time by \citet{2000AJ....120.2938L} using the $F$555$W$ and $F$814$W$ filters of the Advanced Camera Survey (ACS) in NGC\,1023, a Lenticular galaxy. They are defined as a family of red clusters ($V-I>$ 1~mag) similar to GCs, but distinguish from them for having a fuzzy appearance, a fainter magnitude ($M_{\mathrm{V}}$~$>-$7), and a larger effective radius ($\mathrm{R_{eff}}>$7~pc) \citep{2000AJ....120.2938L,2006ApJ...639..838P}. Following their discovery, they were initially thought to be exclusive to Lenticular galaxies \citep[e.g.][]{2002AJ....124.2006F}. These authors suggested that FFs are halo objects similar to GCs, but are formed later than GCs in a star formation event triggered by the accretion of a satellite galaxy. They argued that such a merger event is capable of providing physical conditions for the formation of clusters of diverse sizes and masses. The presence of an actively interacting companion (NGC\,1023A) to NGC\,1023 was taken as evidence of such interacting events in the past. An alternative scenario involving their possible association to the stellar disk populations has been suggested by \cite{2009ApJ...702.1268B,2011A&A...529A.138B}. By carrying out numerical simulations they found that merger of Super Star Clusters (SSCs) can result in long-living extended objects with characteristics similar to that of FFs. Kinematical studies of FFs in NGC\,1023 by \citet{2013A&A...559A..67C} and \citet{2016MNRAS.456.2611C} suggest that they belong to the disk, which effectively rules out the scenario suggested by \citet{2002AJ....124.2006F}.

In recent years, objects sharing the characteristics of FFs, sometimes also referred to as extended clusters, have been reported in the disks of some spiral galaxies (e.g. M51 \citealt{2006ApJ...638L..79H}; M81 \citealt{10.1093/mnras/stt1784}; M101 \citealt{2015ApJ...805..160S}). In spite of the difficulty of detecting these fuzzy objects in the non-uniform disk morphology. \citet{2015ApJ...805..160S} suggested that they could be surviving old disk clusters. 
It is likely that FFs in NGC\,1023 share properties similar to those found in spiral galaxies.

The formation and evolution of FFs remain an open question in the study of stellar populations. Two main scenarios have been proposed to explain their origin. In the first scenario, GCs and FFs are considered distinct populations. GCs form early during galaxy formation \citep{2010ApJ...725.2312O}, whereas FFs emerge later through the merging of compact clusters within cluster complexes. The majority of the compact clusters in these cluster complexes merge within 10 crossing times, forming stable configurations for timescale~$>$~5~Gyr \citep{2002AJ....124.2006F,2009ApJ...702.1268B,2011A&A...529A.138B}. However, an alternative possibility is that FFs are formed through a similar process as GCs, with some GCs expanding over time due to dynamical evolution, to form FFs. During this process, clusters lose mass, becoming fainter, and gradually expand, acquiring a ``fuzzy" appearance \citep[e.g.][]{2010MNRAS.408L..16G}. Determination of age, metallicity and kinematics of a sample of FFs and GCs in this galaxy would be able to firmly establish the nature of its FF population.

Although NGC\,1023 is the galaxy where FFs were first discovered, the only age determination of FFs comes from a stacked spectrum of 11 FFs from \citet{2001AJ....121.2974L}. 
In order to improve this situation, we carried out a study to determine the age, metallicity, and kinematic parameters of the GC and FF candidates in NCG1023 using Multi-Object Spectroscopy (MOS) data from the 10.4-meter Gran Telescopio Canarias (GTC\footnote{Gran Telescopio Canarias is a Spanish-led initiative with contributions from Mexico and the University of Florida, USA. It is located approximately 2300 meters above sea level at the Roque de los Muchachos Observatory on La Palma in the Canary Islands, Spain.}), which is one of the few facilities that is capable of producing good quality spectra of compact objects as faint as 23~mag, the typical magnitude of FFs in this galaxy. NGC\,1023 has a recessional velocity of $v=636 \pm 3 \ \mathrm{km} \ \mathrm{s}^{-1}$  \citep{1999ApJS..121..287H} and is located at a distance of 10.84 Mpc \citep{2016AJ....152...50T}. It presents an almost edge-on orientation (PA\,$\sim 82^{\circ}$) \citep{2003AJ....125..525J}.

The paper is structured as follows. Section \ref{sec:Sec2} describes the sample selection process and details about the spectroscopic observations, data reduction, and determination of the bona fide clusters of the sample. Section \ref{sec:Sec3} focuses on the determination of line-of-sight velocities and explains the procedures for deriving the age and metallicity of GCs and FFs. Section \ref{sec:Sec4} presents an analysis of the determined ages and metallicities. Finally, Section \ref{sec:Sec5} provides the conclusions of the study.

\section{Sample and Observations} \label{sec:Sec2}

New \textit{HST}/ACS images of NGC\,1023 in $F$475$W$ (\textit{g}), $F$814$W$ (\textit{I}) and $F$850$LP$ (\textit{z}) filters covering the entire galaxy have become available after the publication of the FF discovery paper by \cite{2000AJ....120.2938L}. The new observations consist of eight pointings  covering a FoV of $\sim$\,12$\,\times$\,7~arcmin$^{2}$. \cite{2014MNRAS.442.1049F} used a mosaic image formed from all these pointings to characterize the star cluster population over the entire galaxy. They also obtained the effective radius ($\mathrm{R_{eff}}$) for all the red candidates using ISHAPE \citep{1999A&AS..139..393L}. 
Using the definition in \citet{2000AJ....120.2938L}, which distinguishes FFs and GCs based on their $\mathrm{R_{eff}}$, we applied a size cut of $\mathrm{R_{eff}} > 7$~pc to classify the clusters in their sample. This resulted in 81~GCs and 27~FFs.
The MOS instrument has some restrictions that prevent observing all the objects of the sample. Given these restrictions, we could observe 23 GC and 24 FF candidates.
These star clusters span a wide range of radial distances within the galaxy, as shown in Fig.~\ref{fig:HST}, which allows us to examine their characteristics as a function of galactocentric distance.

\subsection{Spectroscopic Observations
}

\begin{deluxetable}{ccccccc}
\tablecolumns{6} 
\tablewidth{\columnwidth}
\tablecaption{\label{tab:logOBNGC1023}Log of the MOS spectroscopic observations with GTC/OSIRIS in NGC\,1023.} 
\tablehead{\colhead{Run}  & \colhead{Date} & \colhead{Time} &  \colhead{AM} & \colhead{Seeing}  & \colhead{Night}\\
& &  (s) &  & ('') & \\
\colhead{(1)} & \colhead{(2)} & \colhead{(3)} &  \colhead{(4)} & \colhead{(5)}  & \colhead{(6)}} 
\startdata
2016B-MOS1 &  2017.01-28 &3$\times$1326 & 1.110 & 0.81 & P\\ 
2016B-MOS2 & 2017-01-28  & 3$\times$1326 & 1.304 & 0.94 & P\\
2016B-MOS3 & 2017-01-28 & 3$\times$1326 & 1.663& 1.10 & P\\
2016B-MOS4 &  2017-01-29 & 3$\times$1326&  1.060& 1.10 & S\\
2016B-MOS5 & 2017-01-31  & 3$\times$1326&1.088  & 1.58 & S\\
2016B-MOS6 & 2017-02-16  & 3$\times$1326& 1.241  & 0.95 & P\\
\enddata 
\tablenotetext{}{\textbf{Notes.} Columns: (1) observing run using the R1000B grism ; (2) observational dates in the format of (year-month-date); (3) exposure time (number of exposures × integration time); (4) mean airmass of the three integrations; (5) seeing measured using stars on the acquisition image; (6) night (P = photometric or S = Spectroscopic).}
\end{deluxetable} 

The spectroscopic data used in this work were taken using the OSIRIS instrument in its MOS mode at the GTC using the Mexican share of the GTC time (proposal ID: GTC6-16BMEX; P.I: Y. D. Mayya). All observations were carried out using the R1000B grism with a slit-width $\sim$\,1.2~arcsec, covering a spectral range from 3700 to 6900~\AA, centered at 5455~\AA\ at a spectral resolution between FWHM\,$\sim 7-8$~\AA. OSIRIS uses two CCDs to cover the FoV in the spatial direction. The observations were carried out using the 2$\times$2 (spatial\,$\times$\,spectral) binning mode, to obtain an effective spatial scale of 0.254~arcsec~pixel$^{-1}$, and a spectral sampling of 2.1~\AA~pixel$^{-1}$. 
The observational campaign consisted of six observing blocks (OBs) taken between January 28$^{\rm th}$ and February 16$^{\rm th}$ of 2017, with the first three blocks taken on the same night one after another, and the remaining three taken on three separate nights. Each observing block consisted of three exposures of 1326 seconds each. Airmass, seeing values and other relevant data are given in Table \ref{tab:logOBNGC1023}. Auxiliary observations of HeAr and Ne arc lamps, as well as a spectroscopic standard star G191-B2B were carried out for each night of observations for the purpose of wavelength and flux calibration, respectively. Additionally, each set of observations consisted of two acquisition images in the Sloan r-band filter, one each with and without the slit-mask, and bias and flat-field images.
OSIRIS/MOS uses a slit-mask plate for positioning the objects on the slits.
The Field of View of OSIRIS is large enough to cover the entire zone where clusters were detected in this galaxy, and hence all OBs used the same pointing with the slitlets being oriented along the East-West direction to be able to maximize the number of objects. This resulted in
an effective exposure time of 6.63 hour on each object. All slitlets are of the same width of 1.2~arcsec, whereas their lengths varied between 7 to 15~arcsec. Overall, we could position 54 slitlets over a FoV of 7.4$^\prime\times2^\prime$, which included 23 GCs, 24 FFs and seven fiducial stars to ensure the accuracy of the pointing. The majority of the slitlets were long enough to allow for the extraction of sky\,$+$\,background spectra. In some cases where it was not possible, we used the sky\,$+$\,background spectrum of the slitlet spatially closest to the slitlet in question. Table \ref{tab:tableCLASS} lists the properties of the observed sample.

\vspace*{15mm}

\subsection{Data Reduction} \label{sec:DataRED}

The spectroscopic data were reduced using {\sc gtcmos}\footnote{ \href{https://www.inaoep.mx/~ydm/gtcmos/gtcmos.html}{ https://www.inaoep.mx/$\sim$ydm/gtcmos/gtcmos.html}}, an {\sc iraf}\footnote{IRAF is an image reduction and analysis program designed by the National Optical Astronomy Observatories.}-based pipeline \citep{2016MNRAS.460.1555G} that works on the files in the GTC raw data directory to reduce OSIRIS/MOS data. The following paragraphs give a summary of the reduction procedure we have followed.

The pipeline starts by joining the two independent CCD images into a single mosaic image, correcting for geometrical distortions. All bias images for a given OB are combined to get a master bias, which is then subtracted from every object exposure. The three object exposures in each OB are then combined using a median algorithm, which in the process eliminates cosmic-ray hits. A master arc image was obtained by summing the arc images of ArHe and Ne. The wavelength calibration was carried out separately for each slitlet, which is then applied to the object image to create a wavelength-calibrated 2D image. 

The pipeline makes use of the standard star to obtain the instrumental response curve for the flux calibration procedure. The same standard star observed on the three nights showed very little variation, because of which a master sensitivity curve suitable for all three nights of observation was prepared by combining all three independent observations.
Atmospheric extinction corrections are applied to both standard stars and targets by combining the extinction curve of the Observatorio de Roque de los Muchachos at La Palma with the recorded air masses of the observations.

\startlongtable
\begin{deluxetable*}{cccrrccccc}
\tablecolumns{10} 
\tablecaption{\label{tab:tableCLASS}Properties of the observed globular and faint fuzzy clusters.} 
\tablehead{
\colhead{Slitlet} & \colhead{RA} & \colhead{DEC} & \colhead{ID} & \colhead{R$_\mathrm{eff}$}  & \colhead{$g_{0}$}  & \colhead{$g_{err}$} & \colhead{$(g-z)_{0}$}  & \colhead{$(g-z)_{err}$} & \colhead{Type}\\
&   \colhead{(deg)} & \colhead{(deg)} & & \colhead{(pc)} & \colhead{(mag)}  & \colhead{(mag)}  & \colhead{(mag)} & \colhead{(mag)}  & \\
\colhead{(1)} & \colhead{(2)} & \colhead{(3)} & \colhead{(4)} & \colhead{(5)} &  \colhead{(6)} & \colhead{(7)}  & \colhead{(8)} & \colhead{(9)} & \colhead{(10)}}
\startdata
1 &  40.1839949 & 39.0821897 & 233 &2.67  &22.420 & 0.014 &  1.408 & 0.017 & GC\\ 
2 & 40.1803845 & 39.0912488 & 361 & 17.26  &23.829 & 0.035 & 0.881 & 0.046 & FF\\
3   & 40.1775724 & 39.0836598 &285 &5.20 &22.167 &0.013 & 1.361 &0.015 & GC\\
4  & 40.1749581 &39.0937148 & 168 & 1.28  & 22.224 &0.014 &  1.146 & 0.016 & GC \\
5  & 40.1708518 & 39.0549741 & 91 & 8.47  &23.508 &0.018 &0.892 & 0.036 & FF\\
7  &  40.1653500 &39.0705794 & 354 & 16.14   & 23.949 &0.036 &1.064 & 0.049 & FF\\
8  & 40.1623983 & 39.0685717 & 222 & 2.27 & 22.176 & 0.013 & 1.015 & 0.017 & GC\\
9 & 40.1597327 & 39.0634583 & 292 & 6.04& 22.957 & 0.021 & 1.083 & 0.027 & GC \\
10 &   40.1568911 & 39.0906317 & 330  & 10.35 & 23.565 & 0.028 & 0.804 & 0.040 & FF \\
11 & 40.1543324 & 39.0910730 & 334 &  11.40& 23.624 & 0.030 & 1.048 & 0.038  & FF \\
12 & 40.1516300 & 39.0713551 & 305  & 7.65 &  23.482 & 0.028  &1.341 &0.036  & FF\\
13 &  40.1489654 & 39.0632847 & 328 & 10.28 & 23.870 & 0.040 & 1.065 & 0.056  & FF\\
14 & 40.1451630 & 39.0900767 & 33 & 3.07& 19.786 & 0.030 & 1.218 & 0.005 & GC \\
15 & 40.1428120 & 39.0670890 & 279  & 4.98& 22.683 & 0.017 &  1.340 &0.022 & GC \\
16 & 40.1401687 & 39.0754275 & 326  & 10.26 & 23.933 & 0.040 & 1.575 & 0.048  & FF\\
17 & 40.1362948 & 39.0506144 & 286 & 5.37& 22.131 & 0.012 & 1.513 & 0.015 & GC \\
18 & 40.1339243 & 39.0740794 & 319 & 9.25& 24.383 & 0.067 & 1.447 & 0.089 & FF \\
19 & 40.1311275 & 39.0685135 & 259 & 3.61& 21.604 & 0.010 & 0.981 & 0.014 & GC \\
21 & 40.1252621 & 39.0725728 & 325 & 10.23 & 23.602 & 0.035 & 1.224 & 0.050 & FF\\
22 & 40.1229300 & 39.0606583 & 345 & 12.97 & 23.037 & 0.024 & 1.477 & 0.032 & FF\\
23 & 40.1201482 & 39.0712822 & 338  & 12.06 & 23.985 & 0.052 &  1.620 & 0.066 & FF\\
24 & 40.1163940 & 39.0771703  & 348  & 13.50 & 23.677 & 0.034 & 1.679 & 0.040 & FF\\
25  & 40.1135061 & 39.0728134 & 375 & 6.44 & 22.793 & 0.026 &  1.423 & 0.035  & GC\\
26 & 40.1117349 & 39.0724285 & 157  & 1.07 & 22.532 & 0.029 & 1.324 & 0.043  & GC\\
27 & 40.1099947 & 39.0720228 & 320  & 9.39 & 23.983 & 0.067 & 1.603 & 0.089  & FF\\
28  & 40.1081361 & 39.0600499 & 364 & 17.60 & 23.605 & 0.096 & 1.626 & 0.128  & FF\\
29 & 40.1022369 & 39.0888212 & 102  & 9.71& 21.186 & 0.005 & 0.987 & 0.009  & FF \\
30 & 40.0997219 & 39.0802891 & 315  & 8.93 & 22.040 & 0.011 &  1.528 & 0.013  & FF\\
31 & 40.0966307 & 39.0659014 & 192  & 1.76 & 23.421 & 0.028  &0.972 & 0.041  & GC \\
33  & 40.0914856 & 39.0757307 & 377 & 9.80 & 23.544 & 0.042 &  1.602 & 0.053  & FF\\
34 & 40.0894504 & 39.0711505 & 356 & 16.49 & 23.660 & 0.041 & 1.384 & 0.056 & FF\\
35 & 40.0873964 & 39.0670821 & 170 & 1.35 & 22.302 & 0.018 & 1.014 & 0.029  & GC\\
36  & 40.0854911 & 39.0591078 & 231 & 2.58 & 22.931 & 0.030 & 0.901 & 0.051  & GC\\
38 & 40.0780876 & 39.0758343 & 18 & 2.12 & 24.159 & 0.045 &  1.185 & 0.098 & GC \\
39 & 40.0748401 & 39.0576972 & 154 & 0.99  & 23.181 & 0.024 & 1.483 & 0.030 & GC\\
40 & 40.0728474 & 39.0628801  & 114 & 12.64& 23.961 & 0.032 & 1.393 & 0.063  & FF \\
41 & 40.0704803 & 39.0642762 & 203  & 1.99& 21.120 & 0.007 &  0.857 & 0.010 & GC \\
42  &  40.0675114 & 39.0577937 & 340 & 12.23  & 23.382 & 0.030 & 1.449 & 0.038 & FF\\
44 & 40.0614984 & 39.0788092 & 289  & 5.58 & 24.913 & 0.093 &  1.622 & 0.116  &  GC\\
45 & 40.0594476 & 39.0540857 & 173 & 1.39 & 22.915 & 0.020 & 0.969 & 0.028  & GC\\
46 & 40.0569413 & 39.0719463 & 336 & 11.63 & 23.980 & 0.040 &  1.381 & 0.050 & FF\\
47 & 40.0544506 & 39.0747080 & 174 & 1.40 & 22.266 & 0.013 &  0.932 & 0.018  & GC\\
48  & 40.0514988 & 39.0467115 & 312 & 8.66 & 23.752 & 0.033 &  0.817 & 0.047  & FF\\
50  & 40.0451795 & 39.0623893 & 160 & 1.12 & 22.377 & 0.014 &  1.228 & 0.018  & GC\\
51 & 40.0421849 & 39.0538877  & 223 & 2.34 & 22.235 & 0.013 & 1.396 & 0.016  & GC\\
53 & 40.0342797 & 39.0538776 & 327 & 10.27 & 23.745 & 0.032  & 1.545 & 0.037 & FF\\
54 & 40.0268752 & 39.0512783 & 152 & 0.90 & 22.223 & 0.013 &  1.211 & 0.016 & GC\\
[1ex]
	\enddata 
	\tablenotetext{}{\textbf{Notes.}  Columns: (1) slitlet number given to the observations in GTC6-16BMEX, the numbers missing are the fiducial stars used for the observation; (2-3) coordinates in degrees in the Equatorial System; (4-9) ID, effective radius, mag, colors and their errors in the \textit{HST} $F$475$W$ (\textit{g}) and $F$850$LP$ (\textit{z}) bands, all from \cite{2014MNRAS.442.1049F}. The 0 subindex indicates that the magnitudes and colors are corrected for the Galactic extinction; (10) objects with R$_\mathrm{eff}>7 \ \textrm{pc}$ are classified as FFs, otherwise it is a GC candidate. }
\end{deluxetable*}

\begin{figure*}[t!]
\includegraphics[width=\linewidth]{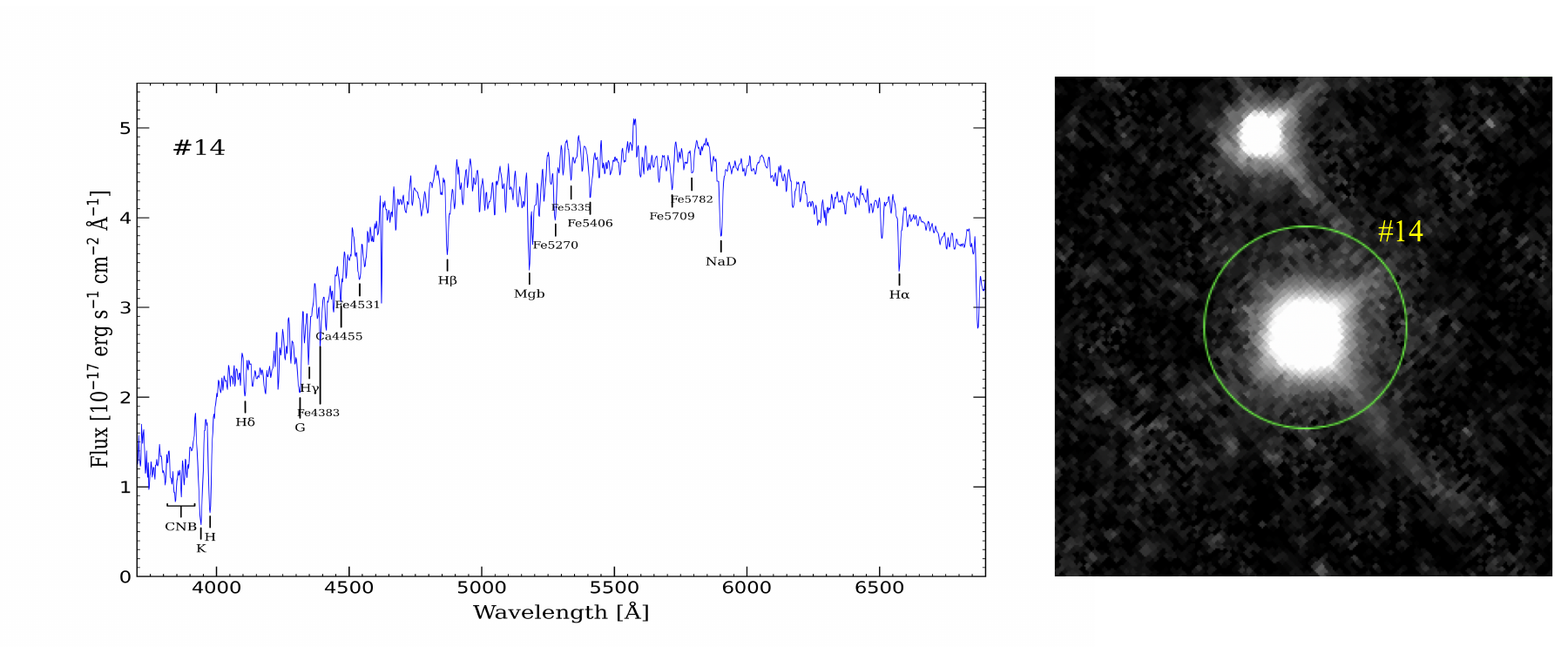}
\caption{OSIRIS reduced spectrum (left) and \textit{HST} grayscale image in the $F$814$W$ filter (right) of candidate in slitlet 14. The \textit{HST} image of the cluster has a size of 5"$\times$5" and an aperture of 1" radius.
\label{fig:GC14_spectra_HST}}
\end{figure*}

The extraction of the spectra is performed interactively using the {\sc iraf} task \textit{apall} on the wavelength and flux-calibrated 2D spectral image. This step leads to an individual 1D spectrum for each target. The apertures of the extraction window of each spectrum were set between 3 and 6 pixels, with the size for each region chosen to maximize the signal-to-noise ratio (SNR) of the extracted spectrum. The chosen aperture size corresponds to the average seeing during observations. For the extraction of sky\,$+$\,background spectra, we used apertures of 4~--~6~pixels in object-free regions of the slitlets.
The trace of the spectrum on the CCD was defined using a Legendre polynomial along the dispersion axis, with the order of the polynomial selected visually to give a best fit. The continuum of the spectra was bright enough (approximately $\mathrm{SNR}>$ 5.0 across the entire spectra) to allow the tracing of the 47 targets. The extraction of all the 47 targets was carried out independently for each of the six OBs. The quality of spectra for OB5 was visibly poor compared to the other five, which was due to relatively poor seeing (1 -- 1.2~arcsec) during this OB. We, therefore did not use the data from this OB and obtained a combined spectrum for each target by co-adding the 15 extracted spectra corresponding to the first four OBs and the last one for an effective exposure time of 3\,$\times$1326\,$\times$\,5~=~5.525~hrs. In columns (3) and (4) of Table \ref{tab:FinalBONAFIDEclusters} and \ref{tab:FinalREJECTDclusters} we give the SNR in the blue and red region of the spectrum between 4500~--~4600 \AA \ and 6400 -- 6500 \AA \, respectively.

\subsection{Separation of false candidates from bona fide clusters}
\label{sec:sepbonadis}

Objects in our spectroscopic sample have been classified as GC or FF candidates based on their R$_\mathrm{eff}$ as measured on the \textit{HST} images. After a visual inspection of their extracted spectra, we noticed that some spectra have spectral characteristics atypical of evolved star clusters, indicating that all candidates are not bona fide clusters. 
We applied the following uniform criteria for an object to be considered as a bona fide cluster:

\begin{enumerate}
    \item The spectrum of the clusters should have a blue (4500 - 4600 \AA) SNR greater than 3.0. This measure ensures sufficient data quality and reliability in the usually faint blue part of the spectrum.
    \item Identification of at least one of the following spectral features in absorption: CNB, G-Band, H$\beta$, MgH, Mg2, Mgb and, Fe (e.g. Fe5270, Fe5335, Fe5406).
    \item  The recessional velocity of the object should be within $\pm$\,400~km\,s$^{-1}$ of the recessional velocity of 636~km\,s$^{-1}$ to the center of the galaxy. This criterion ensures that the selected bona fide clusters are dynamically associated to the host galaxy, and are not compact background sources. 

\end{enumerate}

\begin{deluxetable}{ccrcc}
\tablecolumns{2} 
\tablecaption{\label{tab:FinalBONAFIDEclusters}List of bona fide clusters with their basic spectral characteristics$^{\dagger}$} 
\tablehead{\colhead{Slitlet}   &  \colhead{Type} & \colhead{S/N (blue)}  & \colhead{S/N (red)} & \colhead{Velocity}\\
& & & & (km s$^{-1}$)\\
\colhead{(1)} & \colhead{(2)} & \colhead{(3)} & \colhead{(4)} & \colhead{(5)}  } 
\startdata
&\\[-2.5ex] 
1 & GC & 3.68 & 14.37 & 732.9 $\pm$ 20.8\\
3 & GC & 10.33 & 33.63 & 635.4 $\pm$ 10.5 \\
4 &  GC & 6.69 & 19.72 & 460.3 $\pm$ 15.6\\
9 & GC & 9.09 & 14.32  & 599.1 $\pm$ 50.5 \\
14 & GC & 20.73 & 58.96 & 537.7 $\pm$ 5.30 \\
15 & GC & 8.03 & 22.78 & 816.3 $\pm$ 20.1 \\
16 &  FF & 3.07 & 11.42 &  627.9 $\pm$ 32.7 \\
17 &  GC & 14.38 & 39.12 & 753.9 $\pm$ 9.20 \\
19 & GC & 14.04 & 22.39 & 841.2 $\pm$ 18.1 \\
22 & FF & 7.83 & 15.43 & 753.3 $\pm$ 28.3\\
23 & FF & 3.02 & 5.76 & 720.2 $\pm$ 79.7\\
24 & FF & 18.37 & 16.25 & 736.7 $\pm$ 20.6\\
25 & GC & 9.55 & 16.42 & 616.1 $\pm$ 20.1 \\
26 & GC & 5.67 & 20.45 & 934.4 $\pm$ 25.5\\
27 & FF & 39.52 & 52.29 & 757.4 $\pm$ 5.20\\
28 & FF & 14.18 & 35.72 & 840.8 $\pm$ 15.7 \\
29 & FF & 8.77 & 37.01 & 834.9 $\pm$ 16.1 \\
31 & GC & 3.52 & 11.30 & 602.3 $\pm$ 5.30 \\
35 & GC & 8.34 & 21.43 & 684.8 $\pm$ 39.4 \\
36 &  GC & 7.46 & 9.34 & 509.9 $\pm$ 4.60\\
38 & GC & 10.51 & 27.17 & 638.5 $\pm$ 40.4\\
41 & GC & 17.92 & 33.76 & 695.0 $\pm$ 18.7\\
42 & FF & 5.65 & 8.13 & 556.8 $\pm$ 24.2\\
46 &FF & 3.48 & 7.76 & 641.9 $\pm$ 188.8\\
50 & GC & 22.50 & 63.52 & 769.9 $\pm$ 25.3 \\
51 & GC & 5.20 & 14.80 & 552.6 $\pm$ 40.8\\
54 & GC & 5.33 & 5.44 & 842.9 $\pm$ 16.8\\
[1ex]
\enddata 
\tablenotetext{}{$^{\dagger}$ A visualization of all the bona fide clusters can be found in the Appendix \ref{sec:bonafide} of this work.}
\end{deluxetable}

\begin{deluxetable}{ccccl}
\tablecolumns{5} 
\tablecaption{\label{tab:FinalREJECTDclusters}List of rejected candidates with their basic spectral characteristics} 
\tablehead{\colhead{Slitlet} &  \colhead{Type} & \colhead{S/N (blue)}  & \colhead{S/N (red)} & \colhead{Comments}\\
\colhead{(1)} & \colhead{(2)} & \colhead{(3)} & \colhead{(4)} & \colhead{(5)} } 
\startdata
&\\[-2.5ex] 
2 & FF & 3.66 & 7.76 & Poor SNR \\
5 & FF & 14.20 & 20.97 & Blue cont \\
7 & FF & 4.35 & 9.00 & Blue cont \\
8 & GC & 12.36 & 18.09  & Blue cont\\
10 & FF & 3.47 & 6.48 & BO at \textit{z} = 0.276 \\
11 & FF & 4.39 & 6.36 & BO at \textit{z} = 0.775 \\
12 & FF & 6.75 & 10.22 & BO at \textit{z} = 0.796 \\
13 & FF & 0.43 & 5.18 & Poor SNR \\
18 & FF & 8.25 & 4.51 & Emission spectrum \\
21 & FF & 7.69 & 15.08 & BO at \textit{z} = 0.550 \\
30 & FF & 5.87 & 21.59 & Emission spectrum \\
33 & FF & 1.46 & 8.63 & BO at \textit{z} = 0.135 \\
34 & GC & 2.89 & 14.19 & Poor SNR  \\
39 & GC & 7.84 & 5.45  & Emission spectrum\\
40 & FF & 6.88 & 5.64 & Emission spectrum \\
44 & GC & 7.31 & 6.66 & BO at \textit{z} = 0.445\\
45 & GC & 6.41 & 7.34 & Emission spectrum \\
47 & GC & 6.76 & 27.91 & Emission spectrum \\
48 & FF & 3.14 & -1.87 & Poor SNR \\
53 & FF & -2.52 & 1.15 & Poor SNR\\
[1ex]
\enddata 
\tablenotetext{}{\textbf{Notes.} In the last column, we give the reason for the rejection. The terminology used in this column refers to BO~=~background object, Blue cont~=~spectrum brightens towards blue wavelengths and Emission spectrum~=~ spectra showing emission lines with recessional velocity not coinciding with that expected in NGC\,1023. No detection of absorption features typical of an old stellar population. A visualization of the discarded cluster candidates can be found in the Appendix \ref{sec:discarded} of this work.}
\end{deluxetable}

\vspace*{-10mm}
Twenty of our objects did not meet the three criteria resulting in a sample of 27 bona fide clusters, of which 18 are GCs and nine FFs. The H$\alpha$ in absorption is the most noticeable spectral features in the spectra of bona fide clusters. In Fig. \ref{fig:GC14_spectra_HST}, we show the spectrum of one of the brightest objects of our sample corresponding to slit\#14. As expected many metallic lines typical of old stellar systems are seen in the spectrum. Some of these features drop out in spectra with relatively lower SNR ratio.

Among the 20 targets that did not meet the cluster criteria defined above, six were GC candidates, with the remaining 14 being FF candidates. However, the spectra of five of the 20 candidates are of poor quality (SNR~$<$~3), though they could still potentially be clusters. The remaining 15 can be classified into one of the following types of objects:

\begin{enumerate}
    \item 
    Background sources: Spectra show emission lines consistent with nebular spectra at a higher redshift than NGC\,1023. These are clearly compact background galaxies. Six objects belong to this category.
    \item Featureless spectra: Spectra that are rising towards blue wavelengths. The absence of any identifiable feature makes it impossible to conclude whether they could be objects in NGC\,1023 or background BL Lac kind of Active Galactic Nuclei (AGN). Three objects belong to this category.
    \item Peculiar spectra: Spectra containing emission lines, but not consistent with any known nebular line in the rest-frame UV to optical wavelengths. Six sources are of this kind. 
\end{enumerate}

In Table \ref{tab:FinalBONAFIDEclusters} we list the bona fide clusters along with their recessional velocities measured from our spectra as explained in the next section, and the rejected candidates are listed in Table \ref{tab:FinalREJECTDclusters}, where in the last column, we give the reason for the rejection. We henceforth refer to all bona fide clusters by their slitlet number (e.g.,~GC14).

\section{Analysis} \label{sec:Sec3}

We started our analysis by correcting the spectra of all bona fide clusters for the Galactic extinction in the direction of NGC~1023. For this, 
we used the extinction curve of \cite{1989ApJ...345..245C} (hereafter CCM89), adopting $R_{V}=3.1$ as the value for the diffuse interstellar medium and an extinction value in the V band of ~$A_{V}=~0.166 \ \textrm{mag}$ \citep{2011ApJ...737..103S}.

\vspace{1cm}
\subsection{Line-of-sight velocities} \label{sec:velocity}

In order to obtain the Line-of-sight (LOS) velocities for each bona fide cluster from the spectra, we employed the penalized pixel fitting algorithm \citep[pPXF;][]{2004PASP..116..138C}, which determines stellar population properties by fitting an observed spectrum into a combination of template spectra, allowing us to recover the line-of-sight velocity distribution (LOSVD) of the object. For template spectra, we used the simple stellar populations (SSP) models of \cite{2003MNRAS.344.1000B} hereafter BC03 for the analysis, adopting a Kroupa initial mass function and a fixed metallicity of \textit{Z}~=~0.008. The BC03 models we used correspond to Padova stellar evolutionary tracks \citep{1994ApJS...94...63B} and MILES library of stellar spectra \citep{2006MNRAS.371..703S}. A total of 14 spectra were generated, from 1 to 14~Gyr, in steps of 1~Gyr. 
The \citetalias{2003MNRAS.344.1000B} SSP models were trimmed, resampled and smoothed to the same wavelength coverage, pixel sampling, and resolution of the observed spectra, respectively. The pPXF code fits the observed spectra with the model spectra in pixels of equal logarithmic wavelengths, which transforms the wavelength shifts caused by the Doppler effect into simple linear displacements in pixels. 

For each spectra, two pPXF fits were performed as part of the analysis to obtain a more accurate estimate of the noise. The first pPXF fit is performed using a constant noise. However, the second fit improves the noise determination by using the residuals from the initial pPXF fit. This is achieved through a bootstrapping technique with a fixed interval size in pixels. The interval spans a specific number of pixels and resamples the residuals.
For this technique, we used an interval size of 20 pixels and performed 200 bootstrap iterations. The resampling process involves iteratively selecting random samples with replacements from the residuals within each interval. Then, the noise of the resampled residuals is calculated and assigned to the corresponding indices within each interval. This approach ensures a more robust noise characterization.
The selection of our interval size and the total number of iterations in bootstrapping are based on the previous specific considerations, particularly the need for convergence across iterations. These choices enable us to extract accurate statistics while also optimizing computational time. 
Furthermore, we performed the bootstrapping process for 200 iterations, as changes in the results became negligible and unnoticeable beyond this point ensuring that the bootstrapping procedure has satisfactorily captured the desired statistical characteristics.

The determination of the measured uncertainties of the LOSVD parameters is done by a Monte Carlo simulation set to 500 trials. In each simulation, a new version of the observed spectrum is created by adding Gaussian noise to the spectrum based on the noise calculated as described in the previous paragraph. The median value of these 500 simulations estimates the LOSVD, with the 16th and 84th percentile values taken as lower and upper errors in the measured value. The last column of Table~\ref{tab:FinalBONAFIDEclusters} gives the resulting velocity and its errors of all bona fide clusters. As a final step, each observed spectrum was corrected for redshift using the derived velocity found using the \textit{dopcor} task in {\sc iraf}.

\subsection{Metallicity, Age and Extinction}

Availability of spectra allows the determination of metallicity, age, and extinction of old stellar systems from a detailed analysis of their features that avoids the degeneracy that affects quantities such as colors. This is made possible because of the existence of some spectral features that are more sensitive to age (i.e. H$\beta$\ absorption feature), and some other spectral features that are more sensitive to metallicities (e.g. Fe indices). Extinction does not affect the strength of the spectral features. However, availability of spectra spanning more than 300~\AA\ of wavelength allows the determination of extinction, once the age and metallicities are determined. The technique of fitting the observed spectrum with model spectra to obtain simultaneously the three quantities that is popular in the last few years while analyzing large quantities of spectral data \citep[e.g.][]{2014A&A...569A...1W} does not necessarily take into account the different sensitivities of different features. We here follow the classical method of using the metal-sensitive spectra indices to obtain the metallicity first, followed by determination of age and finally extinction. This method has been used recently by \citet{10.1093/mnras/stae051} to determine the three quantities in a sample of GCs in M81 using spectra obtained with the same spectral setup as the present study.

\subsubsection{Metallicty determination}
\label{sec:metal321}
We adopted two complementary methods to determine the metallicity from our spectra. In the first method, an approximate estimate of age and metallicity was done using the grid method proposed by \cite{2004MNRAS.351L..19T} using a H$\beta$ \ versus [MgFe]$'$ diagram. In the second method, we calculated the strength of the iron Lick indices to calculate the [Fe/H] values of old stellar systems (age~$>$~3~Gyr).

\begin{deluxetable}{cccccc}[ht!]
\tablecolumns{6} 
\tablewidth{\columnwidth}
\tabletypesize{\scriptsize}
\tablecaption{\label{tab:TragerIndexMEASURE}Spectral indices of our clusters in the \citet {1998ApJS..116....1T} system } 
\tablehead{\colhead{Name} & \colhead{H$\beta$} & \colhead{Mgb} & \colhead{Fe52} & \colhead{Fe53} & \colhead{[MgFe]$'$} \vspace{-1mm}\\
& \colhead{(\AA)} & \colhead{(\AA)} & \colhead{(\AA)} & \colhead{(\AA)} & \colhead{(\AA)} \vspace{-1mm}\\
\colhead{(1)}  & \colhead{(2)} & \colhead{(3)} & \colhead{(4)} & \colhead{(5)} & \colhead{(6)}} 
\startdata
&\\[-2.5ex] 
GC1 & 2.33$\pm$0.026 & 2.11$\pm$0.005 & 2.80$\pm$0.009 & 1.90$\pm$0.011 & 2.32$\pm$0.004\\
GC3 & 1.79$\pm$0.002 & 3.92$\pm$0.006 & 2.65$\pm$0.006 & 2.81$\pm$0.007 & 3.25$\pm$0.004\\
GC4 & 2.22$\pm$0.009 & 3.44$\pm$0.007 & 2.65$\pm$0.008 & 0.43$\pm$0.002 & 2.64$\pm$0.005\\
GC9 & 2.45$\pm$0.004 & 1.99$\pm$0.008 & 0.97$\pm$0.007 & 0.72$\pm$0.010 & 1.34$\pm$0.005\\
GC14 & 1.79$\pm$0.005 & 2.66$\pm$0.006 & 2.69$\pm$0.006 & 1.67$\pm$0.007 & 2.53$\pm$0.004\\
GC15 & 1.30$\pm$0.008 & 3.97$\pm$0.007 & 1.18$\pm$0.010 & 1.15$\pm$0.012 & 2.16$\pm$0.008 \\
FF16 & 2.24$\pm$0.004 &4.11$\pm$0.014 & 2.57$\pm$0.002 & 2.24$\pm$0.011 & 3.19$\pm$0.006\\
GC17 & 1.22$\pm$0.006 &2.87$\pm$0.008 & 2.87$\pm$0.007 & 2.89$\pm$0.008 & 2.87$\pm$0.005\\
GC19 & 2.94$\pm$0.005 & 3.08$\pm$0.006 & 1.41$\pm$0.006 & 0.44$\pm$0.003 & 1.87$\pm$0.004\\
FF22 & 2.14$\pm$0.008 & 2.03$\pm$0.012 & 3.44$\pm$0.009 & 3.44$\pm$0.010 & 2.64$\pm$0.009\\
FF23 & 1.96$\pm$0.004 & 0.43$\pm$0.001 & 1.65$\pm$0.002 & 4.94$\pm$0.015 & 1.05$\pm$0.002\\
FF24 & 2.24$\pm$0.003 & 5.15$\pm$0.003 & 2.44$\pm$0.004 & 1.89$\pm$0.006 & 3.43$\pm$0.003\\
GC25 & 1.75$\pm$0.010 & 3.05$\pm$0.009 & 4.66$\pm$0.007 & 4.02$\pm$0.007 & 3.70$\pm$0.006\\
GC26 & 3.63$\pm$0.130 & 3.47$\pm$0.113 & 2.04$\pm$0.112 & 2.62$\pm$0.009 & 2.76$\pm$0.067\\
FF27 & 2.22$\pm$0.005 & 2.51$\pm$0.007 & 2.02$\pm$0.007 & 1.63$\pm$0.009 & 2.19$\pm$0.004\\
FF28 & 1.63$\pm$0.021 & 4.49$\pm$0.013 & 3.41$\pm$0.024 & 2.57$\pm$0.016 & 3.78$\pm$0.012\\
FF29 & 1.93$\pm$0.008 & 0.24$\pm$0.006 & 3.21$\pm$0.007 & 0.51$\pm$0.009 & 0.81$\pm$0.010\\
GC31 & 1.33$\pm$0.006 & 3.04$\pm$0.007 & 1.94$\pm$0.008 & 1.57$\pm$0.005 & 2.36$\pm$0.005\\
GC35 & 1.90$\pm$0.003 & 2.42$\pm$0.003 & 2.22$\pm$ 0.004 & 1.79$\pm$0.003 & 2.25$\pm$0.002\\
GC36 & 1.96$\pm$0.005 & 2.40$\pm$0.004 & 1.68$\pm$0.006 & 1.40$\pm$0.007 & 1.96$\pm$0.003\\
GC38 & 1.59$\pm$0.007 & 4.40$\pm$0.006 & 3.27$\pm$0.007 & 3.75$\pm$0.006 & 3.87$\pm$0.004\\
GC41 & 2.06$\pm$0.004 & 0.71$\pm$0.005 & 2.09$\pm$0.007 & 1.21$\pm$0.008 & 1.14$\pm$0.004\\
FF42 & 3.12$\pm$0.002 & 2.10$\pm$0.005 & 4.02$\pm$0.005 & 4.21$\pm$0.005 & 2.92$\pm$0.004 \\
FF46 &1.51$\pm$0.003 & 3.88$\pm$0.036 & 2.09$\pm$0.003 & 0.74$\pm$0.001 &3.76$\pm$0.018\\
GC50 & 1.60$\pm$0.008 & 0.56$\pm$0.006 & 0.52$\pm$0.007 & 0.68$\pm$0.008 & 0.56$\pm$0.004\\
GC51 & 2.02$\pm$0.010 & 2.91$\pm$0.007 & 3.09$\pm$0.008 & 1.71$\pm$0.012 &2.80$\pm$0.005\\
GC54 & 1.65$\pm$0.248 & 2.42$\pm$0.006 & 5.31$\pm$0.005 & 1.54$\pm$0.008 & 3.21$\pm$0.004\\
[1ex]
\enddata 
\tablenotetext{}{\textbf{Notes.} (1) Cluster name; (2) H$\beta$ \ index; (3) Mgb index; (4) Fe52 index; (5) Fe53 index; (6) [MgFe]$'$ index.}
\end{deluxetable} 

\begin{figure}
\includegraphics[width=\columnwidth]{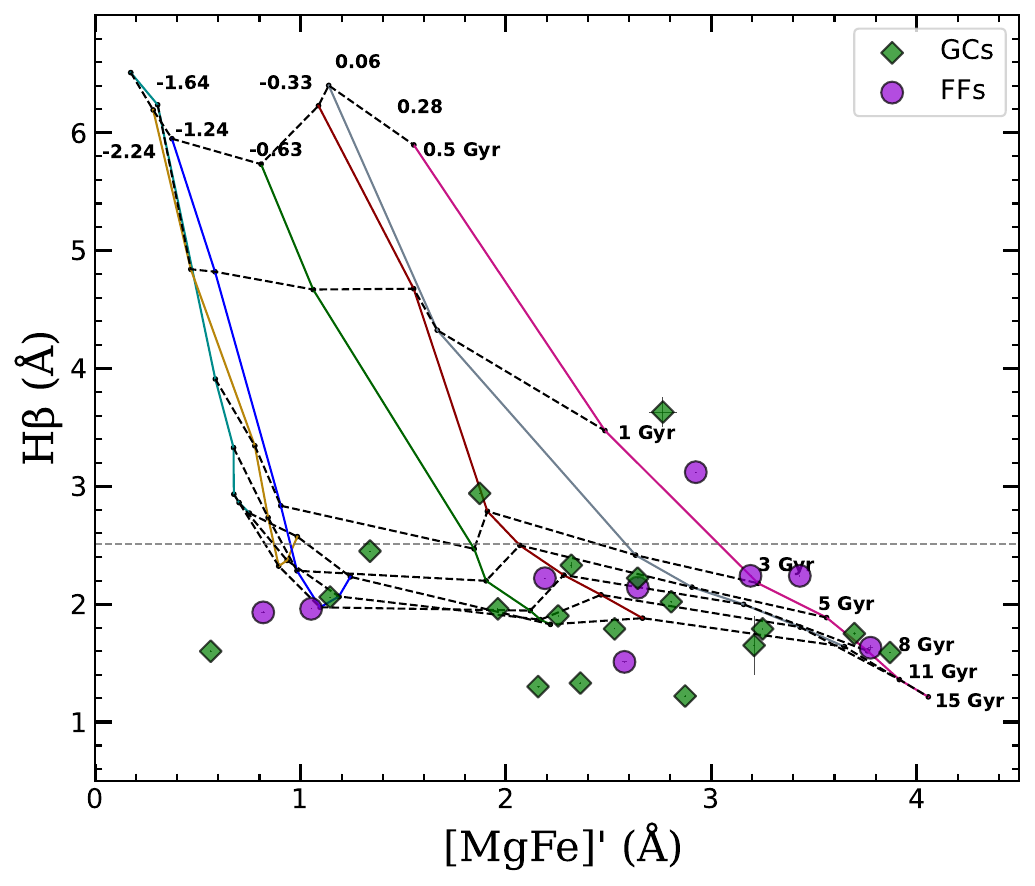}
\caption{H$\beta$ \ vs [MgFe]$'$ grid diagram of the GCs (green solid diamonds) and FFs (violet solid circles). Evolutionary tracks (colored lines) from \citetalias{2003MNRAS.344.1000B} SSPs are shown from 0.5~Gyr to 15~Gyr. Each track corresponds to a fixed [Fe/H] value between $-$2.24 and 0.28. Points on the track that correspond to the same age are connected by the dashed black lines. The horizontal gray dashed line separates relatively younger clusters (above the line) from older clusters.
\label{fig:Tragerindexindex}}
\end{figure}

We start by describing the grid method below. The depth of the absorption features of Mg and Fe are sensitive to metallicities, provided that the systems are old enough so that cool K and M-type stars dominate their integrated light. Systems that are not old enough to have these metallic features are dominated by stars with spectral type G or earlier, which are characterized by the presence of Balmer absorption lines. Hence, the grid method involving H$\beta$\ index versus [MgFe]$'$ index diagram is an excellent diagnostic to break the age-metallicity  degeneracy for systems younger than 3~Gyr. 
The grid method uses the H$\beta$, magnesium (Mgb), and the iron (Fe5270, Fe5335) indices based on the Lick/IDS absorption-line equivalent width (EW)  definitions of \citet{1998ApJS..116....1T}, given by:
\begin{equation}
    \mathrm{EW} =  \int_{\lambda_{2}}^{\lambda_{1}}1- \frac{F_{I}}{F_{BC}+F_{RC}}  \ d\lambda,
\end{equation}
where $F_{I}$ is the mean flux measured in the center of the index, and $F_{B}$ and $F_{C}$ are the mean fluxes in the continuum band passes on the blue and red continuum, respectively. The above equation is used to calculate EWs of H$\beta$ , Mgb, Fe5270, Fe5535 using the wavelength range for the central, blue and red continuum \citep[see Table 2 in][]{1998ApJS..116....1T}.

To create the index-index diagram, we calculated the [MgFe]$'$ index as presented in \cite{10.1046/j.1365-8711.2003.06248.x} where 
\begin{equation}
[\mathrm{MgFe}]'=\sqrt{\mathrm{Mgb}\times\,(0.72\times \mathrm{Fe52} \ + 0.28 \times\mathrm{Fe53})}.
\end{equation}

The [MgFe]$'$ index is a composite index designed to be insensitive to variations in [$\alpha /$Fe] ratios, combining the Mgb and Fe indices, making it an effective tracer of the overall metallicity of stellar populations. The calculated values of the four indices and the composite [MgFe]$'$ index of our clusters, along with their errors, are given in Table \ref{tab:TragerIndexMEASURE}. 
The errors associated with each index are calculated using a Monte Carlo simulation with 1000 iterations. The root mean square error (RMS) is used as the sigma of the Gaussian in these iteration. On the other hand, the error on the [MgFe]$'$ index reported was calculated by propagating the error of the Mgb and Fe indices. 

\begin{deluxetable*}{ccccccccrc}
\tabletypesize{\scriptsize}
\tablecolumns{10} 
\tablecaption{\label{tab:LickIndexMEASURE}Spectral indices of our clusters in the Lick \citeyearpar{1984ApJ...287..586B} system and metallicities.} 
\tablehead{\colhead{Name} & \colhead{[Fe/H]} & \colhead{CNR} & \colhead{G} & \colhead{MgH} & \colhead{Mg2} & \colhead{Mgb} & \colhead{Fe52} & \colhead{Fe53} & \colhead{Fe54} \vspace{-2mm}\\ 
& \colhead{(dex)} & \colhead{(mag)} & \colhead{(mag)} & \colhead{(mag)} & \colhead{(mag)} & \colhead{(mag)} & \colhead{(mag)} & \colhead{(mag)} & \colhead{(mag)} \vspace{-2mm}\\
\colhead{(1)}  & \colhead{(2)} & \colhead{(3)} & \colhead{(4)} & \colhead{(5)} & \colhead{(6)} & \colhead{(7)} & \colhead{(8)} & \colhead{(9)} & \colhead{(10)} } 
\startdata
&\\[-2.5ex] 
GC1 & $-$0.705$\pm$0.442 & - & 0.563$\pm$0.075 & 0.140$\pm$0.019 & 0.368$\pm$0.024 & 0.264$\pm$0.035 & 0.084$\pm$0.010 &  0.023$\pm$0.002 &  0.120$\pm$0.028\\
GC3 & $-$0.233$\pm$0.183 & 0.180$\pm$0.038 & 0.337$\pm$0.048 & 0.081$\pm$0.009 & 0.247$\pm$0.019 & 0.148$\pm$0.026 &  0.073$\pm$0.015 & 0.082$\pm$0.012 & 0.065$\pm$0.019\\
GC4 & $-$0.955$\pm$0.349 & -0.545$\pm$0.014 & 0.139$\pm$0.025 &  0.021$\pm$0.003 & 0.140$\pm$0.018 & 0.131$\pm$0.020 & 0.064$\pm$0.002 & 0.018$\pm$0.004 & 0.077$\pm$0.013\\
GC9 & $-$1.421$\pm$0.373 &  0.100$\pm$0.013 &  0.006$\pm$0.001 &  0.016$\pm$0.001 & 0.137$\pm$0.018 & 0.063$\pm$0.010 & 0.021$\pm$0.002 & 0.020$\pm$0.004 & 0.059$\pm$0.009\\
GC14 & $-$0.556$\pm$0.111 & 0.082$ \pm $0.011 & 0.137$\pm$0.038 & 0.057$\pm$0.012 & 0.157$\pm$0.016  & 0.104$\pm$0.023 & 0.075$\pm$0.015 & 0.043$\pm$0.009 & 0.059$\pm$0.001 \\
GC15 & $-$0.946$\pm$0.457 &  -0.261$\pm$0.040 & -0.017$\pm$0.002 & 0.015$\pm$0.001 & 0.165$\pm$0.020 & 0.153$\pm$0.023 & 0.059$\pm$0.011 & 0.022$\pm$0.001 & 0.066$\pm$0.011\\
FF16 & $-$0.109$\pm$0.087 & 0.041$\pm$0.011 & 0.227$\pm$0.053 & 0.384$\pm$0.024 & 0.373$\pm$0.028 & 0.186$\pm$0.036 &  -0.056$\pm$0.003 & 0.135$\pm$0.028 & 0.133$\pm$0.008\\
GC17 & $-$0.598$\pm$0.468 & 0.240$\pm$0.039 & 0.200$\pm$0.040 & 0.160$\pm$0.014 & 0.350$\pm$0.019 & 0.167$\pm$0.029 & 0.066$\pm$0.010 & 0.042$\pm$0.009 & 0.051$\pm$0.003\\
GC19 & $-$0.3300  & -0.065$\pm$0.030 & 0.058$\pm$0.010 & 0.034$\pm$0.009 & 0.127$\pm$0.016 & 0.111$\pm$0.023 & 0.040$\pm$0.011 & 0.007$\pm$0.001 & 0.017$\pm$0.002 \\
FF22 & 0.185$\pm$0.088 & 0.183$\pm$0.052 & 0.222$\pm$0.066 & 0.154$\pm$0.018 & 0.263$\pm$0.021 & 0.057$\pm$0.017 &  0.110$\pm$0.021 & 0.101$\pm$0.026 &  -0.051$\pm$0.011\\
FF23 & $-$0.622$\pm$0.551& 0.236$\pm$0.039 & -0.428$\pm$0.056 &  0.535$\pm$0.025 & 0.347$\pm$0.027 & -0.027$\pm$0.008 & 0.059$\pm$0.003 & 0.052$\pm$0.004 & -0.028$\pm$0.006 \\
FF24 & $-$0.531$\pm$0.767 & - & -0.293$\pm$0.043 & 0.058$\pm$0.011 & 0.288$\pm$0.015 & 0.272$\pm$0.021 & 0.048$\pm$0.013 & 0.074$\pm$0.017 & -0.084$\pm$0.016 \\
GC25 &  $-$0.383$\pm$0.288   & -0.095$\pm$0.014 & 0.189$\pm$ 0.020& 0.190$\pm$0.017 & 0.300$\pm$0.021 & 0.104$\pm$0.014 & 0.038$\pm$0.001 & 0.110$\pm$0.022 & 0.077$\pm$0.016\\
GC26 & 0.2883 &  0.064$\pm$0.024 & 0.120$\pm$0.041 & 0.235$\pm$0.020 & 0.244$\pm$0.023 & 0.108$\pm$0.013 &  0.068$\pm$0.013 & 0.052$\pm$0.011 & -0.004$\pm$0.001\\
FF27 & $-$0.745$\pm$0.563 &  0.005$\pm$0.002 & 0.115$\pm$0.021 & 0.086$\pm$0.014 & 0.176$\pm$0.018 & 0.091$\pm$0.015 & 0.057$\pm$0.010 & 0.041$\pm$0.003 & 0.037$\pm$0.005 \\
FF28 & $-$0.336$\pm$0.194 & 0.087$\pm$0.013 & 0.340$\pm$0.020 &0.106$\pm$0.042 &0.253$\pm$0.056 & 0.144$\pm$0.008 & 0.089$\pm$0.006 & 0.054$\pm$0.014 & 0.054$\pm$0.016\\
FF29 & $-$1.100$\pm$0.234 & -0.161$\pm$0.037 & 0.235$\pm$0.021 &  -0.021$\pm$0.005 &  0.021$\pm$0.006 & 0.010$\pm$0.001 & 0.090$\pm$0.011 & -0.007$\pm$0.001 & 0.046$\pm$0.014\\
GC31  & $-$0.772$\pm$0.656 & 0.045$\pm$0.012 & 0.135$\pm$0.024 & 0.112$\pm$0.014 & 0.232$\pm$0.019 & 0.109$\pm$0.019 & 0.053$\pm$0.017 & 0.042$\pm$0.011 & 0.043$\pm$0.013\\
GC35 & $-$0.211$\pm$0.478  & -0.230$\pm$0.022 & 0.030$\pm$0.016 & 0.033$\pm$0.010 & 0.043$\pm$0.011 & 0.076$\pm$0.015 & 0.150$\pm$0.014 & 0.027$\pm$0.006 & 0.023$\pm$0.008\\
GC36 & $-$0.932$\pm$0.655 & 0.033$\pm$0.013 & 0.108$\pm$0.019 & 0.065$\pm$0.013 & 0.163$\pm$0.015 & 0.087$\pm$0.021 & 0.044$\pm$0.017 & 0.036$\pm$0.015 & 0.040$\pm$0.011\\
GC38& $-$0.488$\pm$0.293  & - & 0.397$\pm$0.074 & 0.132$\pm$0.015 & 0.207$\pm$0.017 & 0.188$\pm$0.024 & 0.082$\pm$0.001 & 0.044$\pm$0.002 & 0.057$\pm$0.002\\
GC41 & $-$1.121$\pm$0.523 & 0.024$\pm$0.015 & 0.118$\pm$0.027 & -0.047$\pm$0.012 & 0.023$\pm$0.011 &  0.022$\pm$0.002 & 0.048$\pm$0.019 & 0.018$\pm$0.002 & 0.034$\pm$0.008\\
FF42 & 0.2883 & 0.085$\pm$0.018 & 0.119$\pm$0.020 & 0.009$\pm$0.001 & 0.289$\pm$0.015 & 0.065$\pm$0.012 & 0.090$\pm$0.017 & 0.107$\pm$0.009 & 0.086$\pm$0.017\\
FF46& $-$0.522$\pm$ 0.294 & -0.021$\pm$0.001 & 0.165$\pm$0.016 & 0.482$\pm$0.011 & 0.618$\pm$0.014 & 0.250$\pm$0.020 & 0.060$\pm$0.013 & 0.067$\pm$0.012 & 0.053$\pm$0.014\\
GC50 &$-$1.677$\pm$0.359 & 0.019$\pm$0.002 & 0.031$\pm$0.009 & 0.013$\pm$0.005 & 0.033$\pm$0.011 & 0.017$\pm$0.003 & 0.011$\pm$0.003 & 0.015$\pm$0.002 & 0.023$\pm$0.004\\
GC51 & $-$0.144$\pm$0.210   & 0.586$\pm$0.112 & 0.411$\pm$0.097 & 0.158$\pm$0.021& $-$0.036$\pm$0.017 &  0.105$\pm$0.024 & 0.116$\pm$0.025 & -0.058$\pm$0.014 &  0.124$\pm$0.021\\
GC54 & $-$0.955$\pm$0.415 & -0.207$\pm$0.017 &  0.037$\pm$0.012 & 0.066$\pm$0.011& 0.281$\pm$0.016 & 0.089$\pm$0.024 &  0.058$\pm$0.007 &  0.021$\pm$0.002 & 0.165$\pm$0.035\\
[1ex]
\enddata 
\tablenotetext{}{\textbf{Notes.} 1) Cluster name. (2) Measured metallicity. (3) CNR index. (4) G index. (5) MgH index. (6) Mg2 index. (7) Mgb index. (8) Fe52 index. ~(9)~Fe53 index. (10) Fe54 index}
\end{deluxetable*} 

\vspace*{-5mm}

The calculated values for all our clusters are shown in the H$\beta$ \ vs [MgFe]$'$ index diagram in Fig \ref{fig:Tragerindexindex}, where we have also superposed the values of these two indices calculated using the equations~1 and 2 for the \citetalias{2003MNRAS.344.1000B} SSPs for a range of ages and metallicities. 
The figure shows that the majority of the clusters have  H$\beta$~$\leq$~2.5~\AA \ and fall where the SSP ages are older than 7~Gyr.  Only three clusters, two GCs and one FF, have H$\beta$~$\geq$~2.5~\AA \ suggesting that they are younger objects. Two (GC26 and FF46) of these three clusters have [MgFe]$'$ $\geq$~2.5~\AA, and are closest to the highest metallicity SSP model ([Fe/H] = 0.28) that we have plotted. The remaining GC (GC19) lies precisely over the SSP track of [Fe/H]~=~$-$0.33. The majority of clusters have [MgFe]$'$~$\leq$~3~\AA, suggesting [Fe/H] $<-$0.33.

In the second method, we used the empirical calibration between iron Lick indices as defined by \citet{1990ApJ...362..503B} and [Fe/H] to determine exact value of metallicity. The relation was calibrated by \citet{10.1093/mnras/stt1784} using the \cite{2005ApJS..160..163S} library of integrated spectra of Galactic GCs and has been recently been used in \citet{10.1093/mnras/stae051}. To calculate the index, each spectrum is degraded to the resolution specified in \cite{1984ApJ...287..586B}, corrected in redshift and then normalized in flux.
The wavelength limits for each of the three iron indices (Fe5270, Fe5335, Fe5406) are also taken from this work. For the sake of completeness, we also included in the table some other indices such as the redder of the two bands of CN (CNR), the G-band (G4300), the magnesium hydride and the magnesium $b$ triplet (MgH, Mg2, Mgb).

Table \ref{tab:LickIndexMEASURE} displays each measured Lick index along with its corresponding error. The error for each index was determined by performing 1000 Monte Carlo simulations, considering the uncertainties of the fluxes in the center of the index and both continuum band passes.
The [Fe/H] value in Table \ref{tab:LickIndexMEASURE} represents the weighted mean derived from the three iron indices. The determination of the error in the [Fe/H] value involved incorporating the uncertainties of the Fe indices along with the dispersion in the coefficients of the calibration equation \citep[see][]{10.1093/mnras/stae051}

The metallicity obtained from this second method is not reliable for systems younger than $\sim$3~Gyr, as the sample of Galactic GCs against which this method was calibrated does not include such young systems. We hence assigned the approximate metallicities obtained from the grid method for the three relatively young clusters (GC19, GC26 and FF42).

\subsubsection{Spectroscopic ages and extinction}\label{sec:spectroscopicages}

The grid method described above has given us an approximate value of the age. However, this approach is not used for the final determination of ages in this work. Rather, it is used mainly to distinguish between young and old clusters, offering a quick classification to identify young clusters (H$\beta\geq 2.5$ \AA). We here describe the technique we have followed to get a more accurate age, using the full spectral fitting. For this, we employed a $\chi^{2}$ fitting method between our Doppler-corrected observed spectra over the entire wavelength range of 3700 \AA \ to 6900~\AA, and SSP model spectra covering ages from 1 to 14 Gyr in steps of 0.5~Gyr.~The \citetalias{2003MNRAS.344.1000B} SSP models with a Kroupa initial mass function, stellar evolutionary tracks from PARSEC \citep{2012MNRAS.427..127B} and stellar spectra from the MILES library \citep{2006MNRAS.371..703S} were used. The model metallicity was fixed at one of the values among $Z$~=~0.0004, 0.004, 0.008, 0.02, and 0.05, that is nearest to the determined metallicity value.

\begin{deluxetable*}{crrccrc}
\tablecaption{\label{tab:besfittingvalues} Best-fit values} 
\tablecolumns{7} 
\tabletypesize{\small}
\tablehead{\colhead{Cluster} &\colhead{[Fe/H] } & \colhead{Age} & \colhead{ $A_{v}$} & \colhead{$\chi^{2}$} &  \colhead{$M_{g0}$} & \colhead{log $\left( \mathrm{M_{cl}}/ \mathrm{M}_{\odot}\right )$} \vspace{-2mm}\\
\colhead{} & \colhead{(dex)} & \colhead{(Gyr)} & \colhead{(mag)} & \colhead{} & \colhead{(mag)} & \colhead{} \vspace{-2mm}\\
\colhead{(1)} & \colhead{(2)} & \colhead{(3)} & \colhead{(4)} & \colhead{(5)} & \colhead{(6)} & \colhead{(7)}}
\startdata
GC1 & $-$0.71$\pm$0.44 & $11.00\pm2.00$ & 0.48$\pm$0.10& 1.71 & $-$8.75$\pm$0.014 & 6.12$\pm$0.08\\
GC3 & $-$0.23$\pm$0.18  & $10.50\pm1.50$ & 0.15$\pm$0.01 & 1.66 & $-$8.51$\pm$0.013 & 6.01$\pm$0.08 \\
GC4  & $-$0.96$\pm$0.35 & $10.00\pm1.50$&  0.60$\pm$0.03 & 1.65 & $-$8.05$\pm$0.014 &5.81$\pm$0.08\\
GC9  & $-$1.42$\pm$0.37 &  $8.00\pm2.00$ & 0.31$\pm$0.08 & 1.83 & $-$9.22$\pm$0.021 & 6.19$\pm$0.11 \\
GC14 & $-$0.56$\pm$0.11 & $10.00\pm1.00$ & 0.55$\pm$0.04 & 0.98 & $-$10.39$\pm$0.030 & 6.75$\pm$0.17  \\
GC15 & $-$0.95$\pm$0.46 & $11.00\pm1.50$ & 0.27$\pm$0.05 & 1.81 & $-$7.99$\pm$0.017 & 5.82$\pm$0.09 \\
FF16 & $-$0.11$\pm$0.09 & $8.00\pm2.00$&0.73$\pm$0.21 & 1.76 & $-$8.74$\pm$0.040 & 6.00$\pm$0.22 \\
GC17 &$-$0.60$\pm$0.47 &  $11.50\pm1.50$  & 0.83$\pm$0.14 & 1.55 & $-$8.54$\pm$0.012 & 6.04$\pm$0.06 \\
GC19  & $-$0.33\rlap{$^{\alpha}$} & $3.00\pm1.50$ & 0.68$\pm$0.25& 1.78 & $-$8.57$\pm$0.010 & 5.57$\pm$0.05\\
FF22  & 0.19$\pm$0.09 & $9.00\pm2.00$& 0.54$\pm$0.09& 1.86 & $-$8.14$\pm$0.024 & 5.80$\pm$0.13 \\
FF23  & $-$0.62$\pm$0.55 & $7.50\pm2.50$ & 0.63$\pm$0.11& 1.80 & $-$6.69$\pm$0.052 & 5.16$\pm$0.29 \\
FF24  &$-$0.53$\pm$0.78& $7.00\pm2.00$ & 0.61$\pm$0.16&  2.18  & $-$6.50$\pm$0.034 & 5.06$\pm$0.19 \\
GC25  & $-$0.38$\pm$0.29 &  $11.00\pm2.00$ & 0.93$\pm$0.07 & 1.51  & $-$8.38$\pm$0.026 &5.97$\pm$0.15 \\
GC26 & 0.28\rlap{$^{\alpha}$} & $3.00\pm1.50$ & 1.16$\pm$0.14 & 1.84 & $-$8.64$\pm$0.029 & 5.45$\pm$0.16 \\
FF27  & $-$0.75$\pm$0.56 & $8.00\pm1.00$ & 0.39$\pm$0.02 & 1.54 & $-$6.19$\pm$0.067 & 4.98$\pm$0.38\\
FF28 & $-$0.34$\pm$0.19& $8.50\pm1.00$\rlap{$^{\dagger}$} & 0.91$\pm$0.17 & 1.24 & $-$8.07$\pm$0.096 & 5.76$\pm$0.54\\
FF29  & $-$1.10$\pm$0.23 & $8.50\pm1.50$\rlap{$^{\dagger}$} &  0.56$\pm$0.10& 1.18  & $-$9.49$\pm$0.005 & 6.32$\pm$0.03  \\
GC31 &  $-$0.77$\pm$0.66 & $10.50\pm1.00$ & 0.48$\pm$0.06& 1.39 & $-$6.80$\pm$0.028 & 5.32$\pm$0.15 \\
GC35  & $-$0.21$\pm$0.48  & $11.00\pm2.50$& 0.29$\pm$0.03& 1.85 & $-$7.87$\pm$0.018 & 5.77$\pm$0.10  \\
GC36 & $-$0.93$\pm$0.65 & $12.00\pm1.00$  & 0.66$\pm$0.05 & 1.17 & $-$7.74$\pm$0.030 & 5.75$\pm$0.17 \\
GC38  & $-$0.49$\pm$0.29 & $8.00\pm2.00$ & 0.74$\pm$0.15 &  1.84 & $-$6.52$\pm$0.045 & 5.11$\pm$0.25 \\
GC41  & $-$1.12$\pm$0.52 & $10.00\pm1.50$ & 0.58$\pm$0.06 & 1.88  & $-$9.06$\pm$0.007 & 6.21$\pm$0.04 \\
FF42 & 0.28\rlap{$^{\alpha}$} & $7.50\pm2.00$\rlap{$^{\dagger}$} & 0.44$\pm$0.04 & 2.09 & $-$7.29$\pm$0.030 & 5.40$\pm$0.17  \\
FF46 & $-$0.52$\pm$ 0.29 & $8.50\pm3.50$ & 0.60$\pm$0.11& 2.63  & $-$6.70$\pm$0.040 & 5.19$\pm$0.23 \\
GC50 & $-$1.68$\pm$0.36 & $11.00\pm1.50$ & 1.09$\pm$0.11& 1.74 & $-$7.80$\pm$0.014 & 5.74$\pm$0.08 \\
GC51  & $-$0.14$\pm$0.21 & $11.50\pm2.00$  &0.65$\pm$0.13 & 2.84 & $-$9.44$\pm$0.013 & 6.42$\pm$0.08 \\
GC54  & $-$0.96$\pm$0.41 & $12.50\pm2.50$\rlap{$^{\dagger}$} & 0.40$\pm$0.21  & 2.90 & $-$8.45$\pm$0.013 & 6.05$\pm$0.07\\
[1ex]
	\enddata
\tablenotetext{}{\textbf{Notes.} (1) Cluster name. (2) [Fe/H] value derived from the spectral indices. The values denoted by $^{\alpha}$ correspond to the metallicity obtained by the grid method. (3) Age and its associated error found through Monte Carlo simulation. The ages denoted by a $^{\dagger}$ symbol represent those clusters with a possible blue horizontal component (see Sec.~\ref{sec:bhb}). (4) Visual extinction. (5) Best $\chi^{2}$ value. (6) Reddening-corrected absolute magnitude. (7) Photometric mass. }
\end{deluxetable*}

\vspace*{-5mm}

As in Sec.~\ref{sec:velocity}, the SSP spectra were trimmed, resampled and smoothed to match the wavelength coverage, sampling, and spectral resolution of the Doppler-corrected observed spectra. Following this, both the observed and model spectra were normalized in flux by dividing them by the mean flux in a windows of 20~\AA\ centered at 5000~\AA. Once the metallicity is fixed to the previously determined value, the shape (i.e. the color) of the observed spectrum is dictated by age and extinction, whereas the strength of the spectral features is dictated by its age only. Thus, it is possible to avoid the degeneracy between the age and extinction if we analyze ranges of wavelengths that are large enough to contain age-sensitive stellar spectral features, but small enough not to be affected by extinction. We found that wavelength ranges of $\Delta\lambda$ = 300~\AA\ is a good compromise. The effect of extinction in each of these relatively small segments can be ignored as a first approximation. At the same time, we ensured that each fitted segment contains enough spectral features that are age-sensitive. For this purpose we did not use the spectra long ward of 5500~\AA, as this part does not have detectable spectral features at the 8~\AA\ resolution of our observations. We defined a weighting factor, $w_{k}$, for each segment $k$ using the model spectra, which is a good indicator of the prevalence of spectral features in the segment. The $w_{k}$ for each model spectrum is defined as
\begin{equation}
    w_{k}= \frac{\sigma^{2}_{k} (\mathrm{Mod})}{\sum_{k=1}^{n_{k}} \sigma^{2}_{k} (\mathrm{Mod})},
\end{equation}
where $\sigma^{2}_{k} (\mathrm{Mod})$ represents the RMS of the SSP model spectra in the $k^{th}$ window. This RMS of the model spectrum is calculated over the entire $\Delta\lambda$ = 300~\AA\ extent of the segment $k$. Segments with larger $\sigma_{k}$, and hence larger $w_k$, contain more prominent spectral features.

\begin{figure}
\includegraphics[width=\columnwidth]{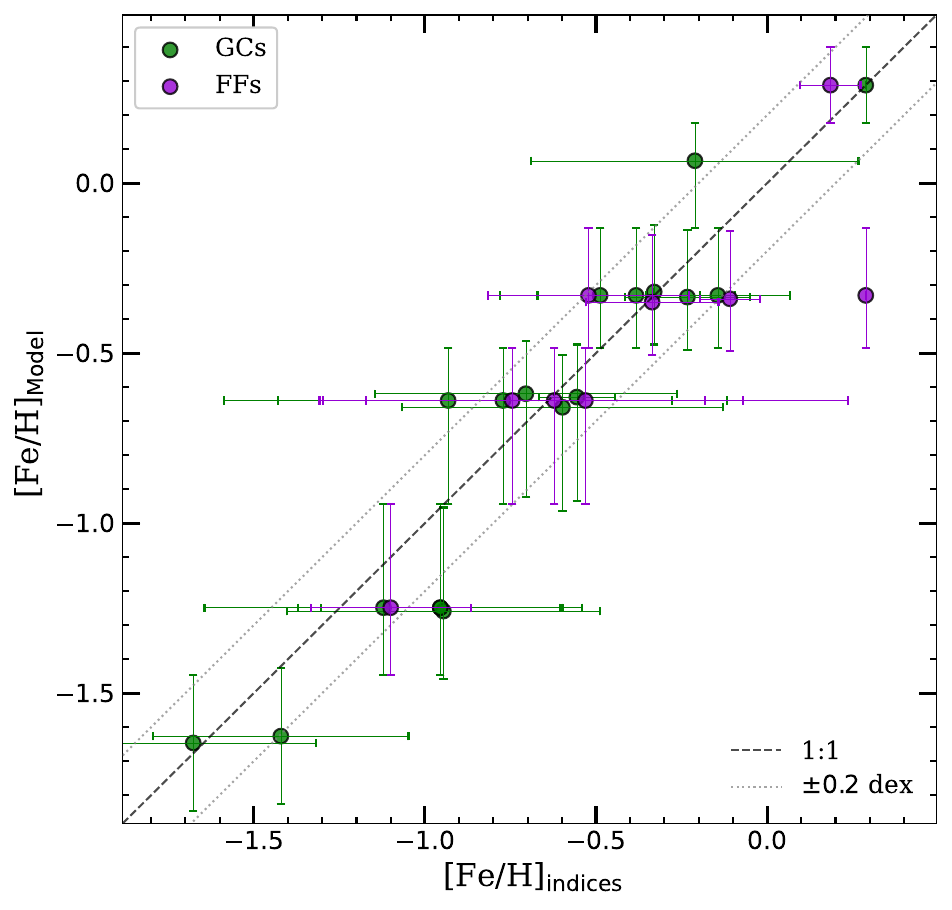}
\caption{Comparison between metallicity determination derived from full spectral fitting and those obtained using spectral indices. Each point represents an individual cluster: GCs (green solid diamonds) and FFs (violet solid circles). The solid line indicates the one-to-one relation, while the dashed lines indicate a 0.2 dex scatter. Small offsets were applied to overlapping points with identical [Fe/H]$_{\mathrm{Model}}$ values for an optimal visualization}.
\label{fig:metalcomparison}
\end{figure}

$\chi_{k}^{2}$ for the $k^{th}$ window is calculated using the expression:
\begin{equation}\label{eq:chi2k}
\chi_{k}^{2}= \frac{1}{n_{k}-1} \sum^{n_{k}}_{i=1} \left (  \frac{\mathrm{\widetilde{Obs}}_{\mathrm{}k}(\lambda_{i})-\mathrm{\widetilde{Mod}}_{\mathrm{}k}(\lambda_{i})}{\widetilde{\sigma_{\mathrm{obs,}i}}}\right )^{2}, 
\end{equation}
where $n_{k}=N-m_{k}$ is the number of unmasked pixels, $N$, the total number of pixels, and $m_{k}$, represents the number of masked pixels, all in the $k^{th}$ window. Masks are designed to ignore pixels affected by strong skylines (e.g. [OI]$\lambda$5577 \AA \ and [OI]$\lambda$6300 \AA), and pixels in the end parts of the spectrum. The normalized observed flux at a specific window is represented by $\mathrm{\widetilde{Obs}}_{\mathrm{}k}$, while
$\mathrm{\widetilde{Mod}}_{\mathrm{}k}$ denotes the normalized model spectrum at the same window.
To apply Eq. \ref{eq:chi2k}, we first calculated the value of $\sigma_{\mathrm{obs,}i}$ defined as
\begin{equation}
\sigma_{\mathrm{obs},i}= \sqrt{\frac{1}{m-1} \sum_{i=1}^m \left ( \mathrm{Obs},i -\mathrm{\overline{Obs}\,} \right ) ^{2}}. 
\end{equation}
Here, the value of $\sigma_{\mathrm{obs,}i}$ is the RMS at each pixel of the observed spectrum. The RMS is calculated over windows of $m$ = 21 unmasked pixels centered on each pixel through an iterative process, clipping pixels that lie outside 2\,$\times$\,RMS from the mean in each iteration. 

An age from the $k^{th}$ window of the spectrum is obtained by the minimization of $\chi_{k}^{2}$. We used the value of age in the window for which $w_{k}$ is maximum as the final best-fit age for the cluster. 
To quantify the uncertainties in the age, we perform an MC simulation with 500 iterations per window. In each iteration, a Gaussian noise with sigma~=~RMS of the spectrum is added to the observed spectrum. The fitting procedure is repeated, and the resulting age distribution is used to determine the uncertainty, with the spread taken as the associated uncertainty. The final $\chi^{2}$ value is obtained as a weighted mean of all $\chi^{2}_{k}$ via
\begin{equation}
    \chi^{2}= \sum_{k=1}^{n_{k}} \ \chi_{k}^{2} \,w_{k}.
\end{equation}
In addition to the metallicity results presented in Sec.~\ref{sec:metal321}, we derived metallicity values simultaneously with age by identifying the best-fitting model, that is, the combination of age and $Z$ model that minimized the $\chi^{2}$ value in each iteration of the MC simulation, thereby obtaining the metallicity associated with the fit. The results are shown in Fig.~\ref{fig:metalcomparison}, where we compare the metallicities obtained from spectral fitting to those derived from spectral indices. Overall, there is good agreement between the methods for most clusters. The main difference lies in the nature of the outputs: spectral fitting yields metallicities that are limited to the discrete values defined by the input models, while the spectral indices provide values that are independent of the model and are not constrained by model steps.
In the determination of age from each segment of the spectra, we have ignored the effect of extinction as mentioned above. However, we obtain the value of extinction and a refined age in an iterative way, as explained below. Although over the $\Delta\lambda$ = 300~\AA\ range, reddening of the spectra due to extinction effects cannot be perceived it is easily noticeable when the model spectrum for the best-fit age is compared with the observed spectrum over the entire wavelength range of our observations. We used this difference in slopes between the observed and best-fit model spectrum to determine $A_V$. Specifically, $A_{V}$ was determined by comparing the observed flux at two specific wavelengths: the blue side of the spectrum at $\lambda_{B}$ = 4400~\AA\ and the red side of the spectrum at $\lambda_{R}$ = 6400~\AA, and using  the following equation:
\begin{equation} \label{eq:extinction}
A_{V}^{B,R} =\frac{\log (F^{B,R}_{\mathrm{obs}}) - \log (F^{B,R}_{\mathrm{mod}})}{-0.4[E_{\mathrm{MW}}(\lambda^{B,R})-1]},
\end{equation}
where $E_{\mathrm{MW}}$ is the Galactic extinction curve of CCM89 evaluated at $\lambda_B$ and $\lambda_R$. The terms $F^{B,R}_{\mathrm{obs}}$ and $F^{B,R}_{\mathrm{mod}}$ represent the average flux values over approximately 200 \AA \ windows centered at $\lambda_B$ and $\lambda_R$, measured for the blue and red regions of the observed spectrum and the best-fitting model spectrum, respectively, with the observed and model spectra normalized by the flux in the $V$-band. Once the $A_{V}$ values are obtained for both the blue and red regions, the mean value is calculated to represent the total extinction of the spectrum. After calculating the mean $A_{V}^{B,R}$ and its associated error, all the model spectra were reddened, and the whole $\chi^{2}$ fitting procedure is repeated to refine the fit between the observed and model spectra. The fit for one of the clusters is shown in Fig. \ref{fig:agebestfit}. The best-fit spectra for the remaining confirmed clusters are provided in the Appendix \ref{sec:bonafide}.

\begin{figure}
\includegraphics[width=\columnwidth]{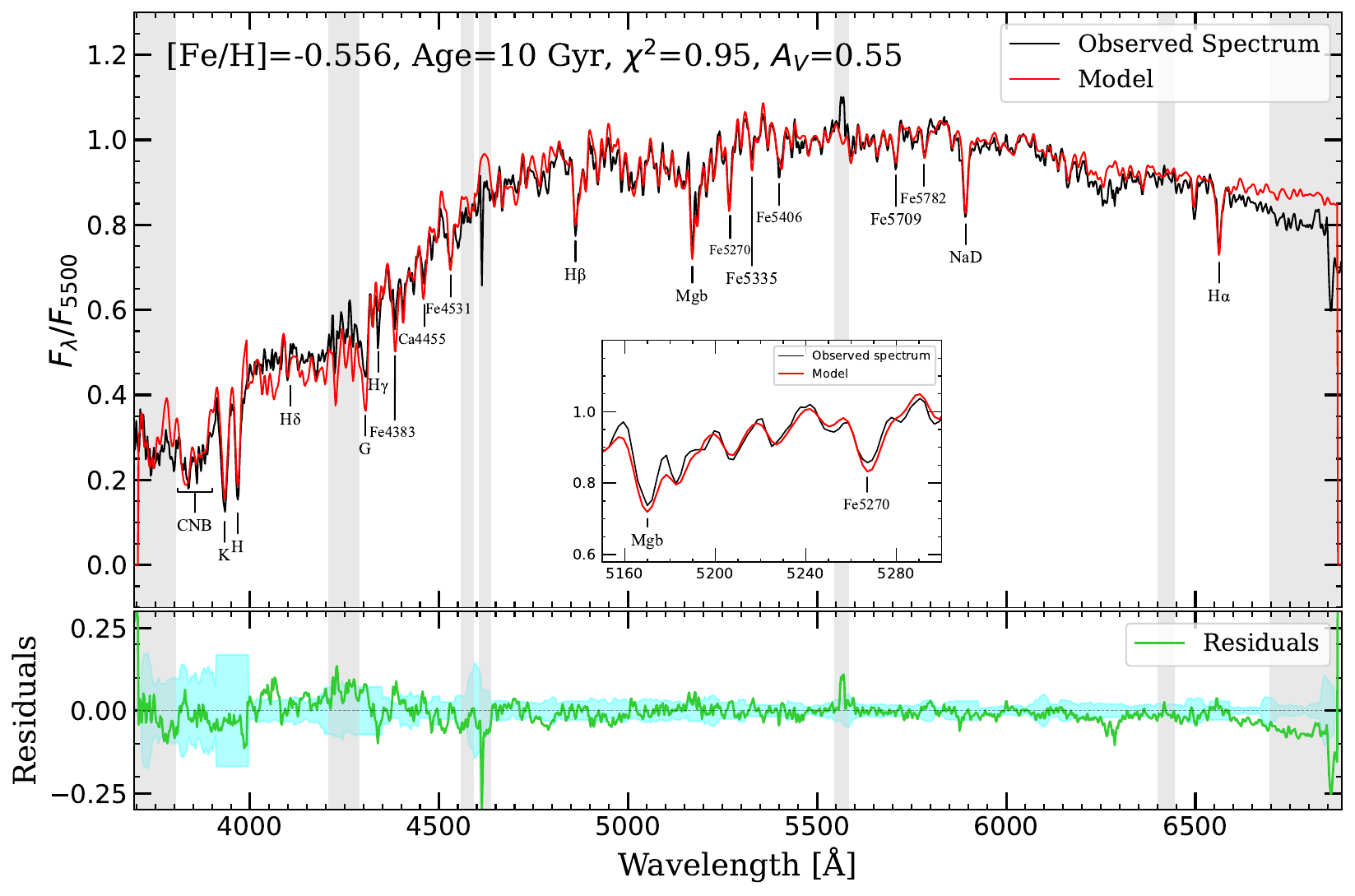}
\caption{Results of the fitting to obtain the age for the illustrative case of GC14. The observed and best-fit model spectra are shown in black and red colors, respectively. A part of the spectrum that includes the Mgb and Fe5270 line is shown zoomed in the inset. The gray vertical bands correspond to the masked wavelengths in the fitting. The residual (obs-model) of the fit is shown in the bottom of the plot, where the cyan horizontal band corresponds to the 1-sigma error. It can be seen that the residuals are less than 1-sigma error at the majority of the unmasked wavelengths.
\label{fig:agebestfit}}
\end{figure}

The results of the best-fit values for age, $A_{V}$ and $\chi^{2}$ are summarized in Table \ref{tab:besfittingvalues}. 
The uncertainties in the age parameter are calculated based on the range of the minimum and maximum ages obtained in the analyzed windows.

\begin{figure*}
\includegraphics[width=\textwidth]{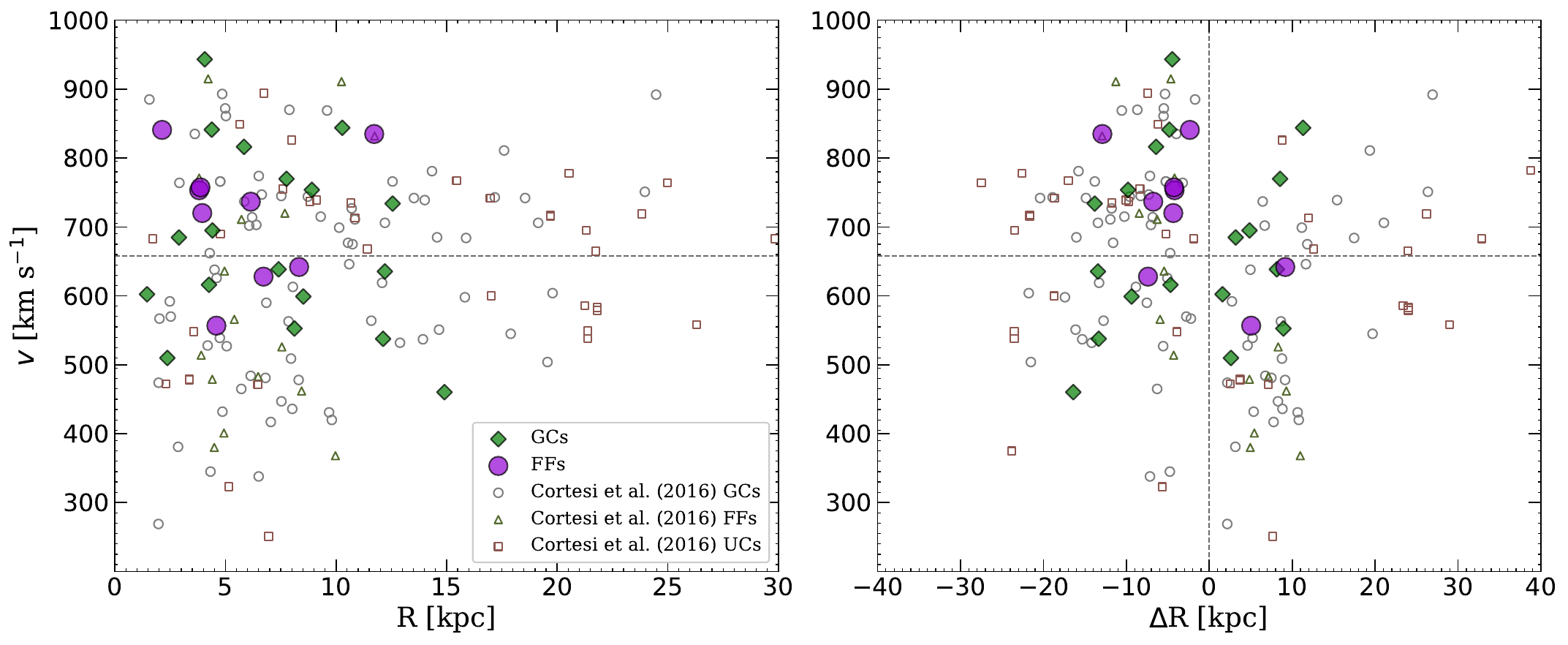}
\caption{The position-velocity maps for GCs and FFs in NGC\,1023. Objects from our study are shown by large symbols as compared to those from the study of \citet{2016MNRAS.456.2611C} (see the inset box in the left-panel for identification of the symbols. UCs stand for unclassified clusters.). The systemic velocity is shown by the horizontal gray line. The x-axis on the left panel is the galactocentric distance, and the right panel, the distance along the major axis, with the negative axis representing the eastern side.
\label{fig:Cortesiversusours}}
\end{figure*}

Lastly, the photometric mass was determined using the following equation:
\begin{equation}
\frac{\mathrm{M_{cl}}}{\mathrm{M}_{\odot}} = 10^{\left(-0.4 ( M_{g0}-M_{g^{\, \mathrm{SSP}}}(t) )\right)},
\end{equation}
where $M_{g0}$ represents the absolute magnitude in the $g$ filter and corrected for reddening using the calculated $A_{V}$ (see column 4 of Table \ref{tab:besfittingvalues}). $  {M_{g^{\mathrm{SSP}}}}(t)$ indicates the $g$ magnitude predicted by the \citetalias{2003MNRAS.344.1000B} SSP models for a cluster with a total mass of 1 ${\mathrm{M}_{\odot}}$ at a determined age. The mass of each cluster is given in the last column of Table \ref{tab:besfittingvalues}.

For completeness, we also determined the internal extinction of our clusters using the color distribution, as the typical reddening experienced by these clusters. Specifically, we derived $A_{V}\mathrm{(int)}$ from the observed $(g-z)_{0}$ color (see Table~\ref{tab:tableCLASS}) by comparing it to the intrinsic color a SSP model with an age of 13.5 Gyr and metallicity $Z = 0.001$, representative of typical stellar populations. The extinction was calculated for each cluster using, $A_{V}\mathrm{(int)} = (g-z)_{0}-(g-z)_{\mathrm{SSP}}/ (E_{g}- E{z})$,
where $E_{g}$ and $E_{z}$ are the values from the \citet{1989ApJ...345..245C} extinction curve at the effective wavelengths of the $g$ and $z$ bands, respectively. This method gives a mean value of $A_{V}\mathrm{(int)} = 0.54 \pm 0.05$ mag, which is consistent with the mean extinction value of $A_{V} = 0.60 \pm 0.10$ mag obtained using the spectroscopic method using Eq.~\ref{eq:extinction}. The $A_{V}\mathrm{(int)}$ represents a lower limit for extinction, since assuming a fixed old and metal-poor SSP does not account for the possibility of younger or more metal-poor clusters.

\section{Discussion} \label{sec:Sec4}

Among the 47 star cluster candidates selected for the spectroscopic study, 20 were excluded from further analysis, of which 12 clearly showed spectral features inconsistent with the spectrum of an old system in NGC\,1023. Specifically, the cluster candidates in slitlets 10, 11, and 12 showed emission lines consistent with redshift values of $z=$ 0.276, $z=$ 0.775, and $z=$ 0.796, respectively. Similarly, the cluster candidate in slitlet 21 exhibited a redshift value of $z=$ 0.550, as measured from the intense [O II]$\lambda$3727 and Balmer lines in emission. Furthermore, the cluster candidate in slitlet 40 was identified as a Galactic object rather than a star cluster associated with NGC\,1023 due to the low measured radial velocity from its emission lines. 
However, for the remaining seven cluster candidates, a definitive rejection was not possible as they were excluded from our study due to insufficient SNR. These cluster candidates lacked absorption or emission features that can be identified with any known features. Overall, the majority (15) of the rejected star clusters are FF candidates, illustrating the necessity for spectroscopic observations to determine their nature as star clusters in the host galaxy.

In the rest of this section, we discuss the kinematical, age, and metallicity results obtained from the 27 bona fide clusters, including the implications of these results regarding the origin of FFs.

\subsection{Kinematical properties of GCs and FFs in NGC\,1023} \label{sec:4.1}

Ten of our 27 clusters with velocity measurements were studied in two previous works, where a total of 135 clusters were analyzed kinematically \citep{2013A&A...559A..67C,2016MNRAS.456.2611C}. These studies found that the GCs belong to the spheroidal component, whereas the FFs are disk objects.

In Fig.~\ref{fig:Cortesiversusours}, we show the velocities of our 27 clusters plotted against the galactocentric distance. The velocities for the sample of \cite{2016MNRAS.456.2611C} are also shown. This latter sample includes some objects (UCs) that could not be cross-matched with the list of \cite{2014MNRAS.442.1049F} due to lack of precision in the coordinates in the latter list. Our velocities are well within the range of velocities reported by \citet{2016MNRAS.456.2611C}, who using their much larger sample concluded that the FFs belong to the disk and the GCs belong to the spheroidal component.

 \begin{figure}[t!]
\includegraphics[width=\columnwidth]{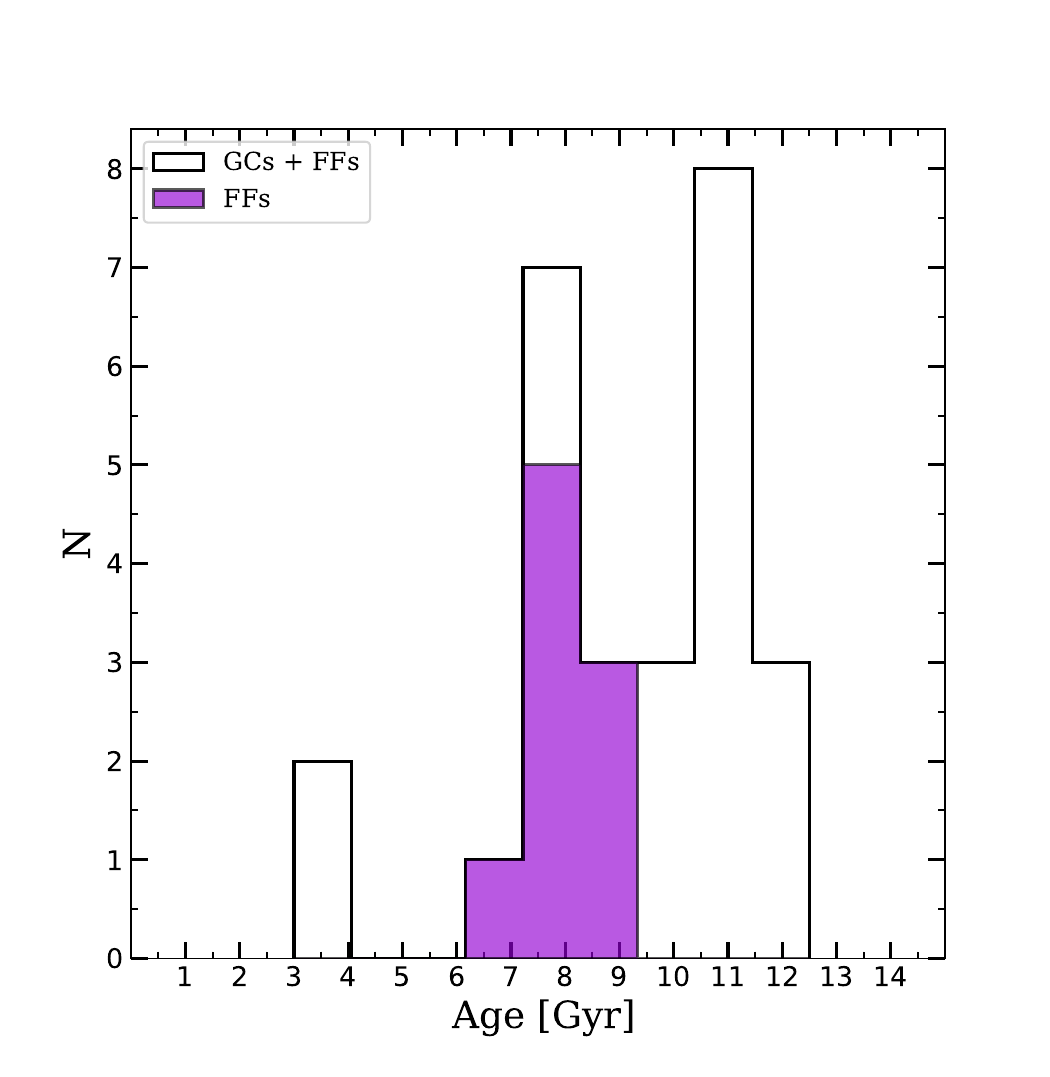}
\includegraphics[width=\columnwidth]{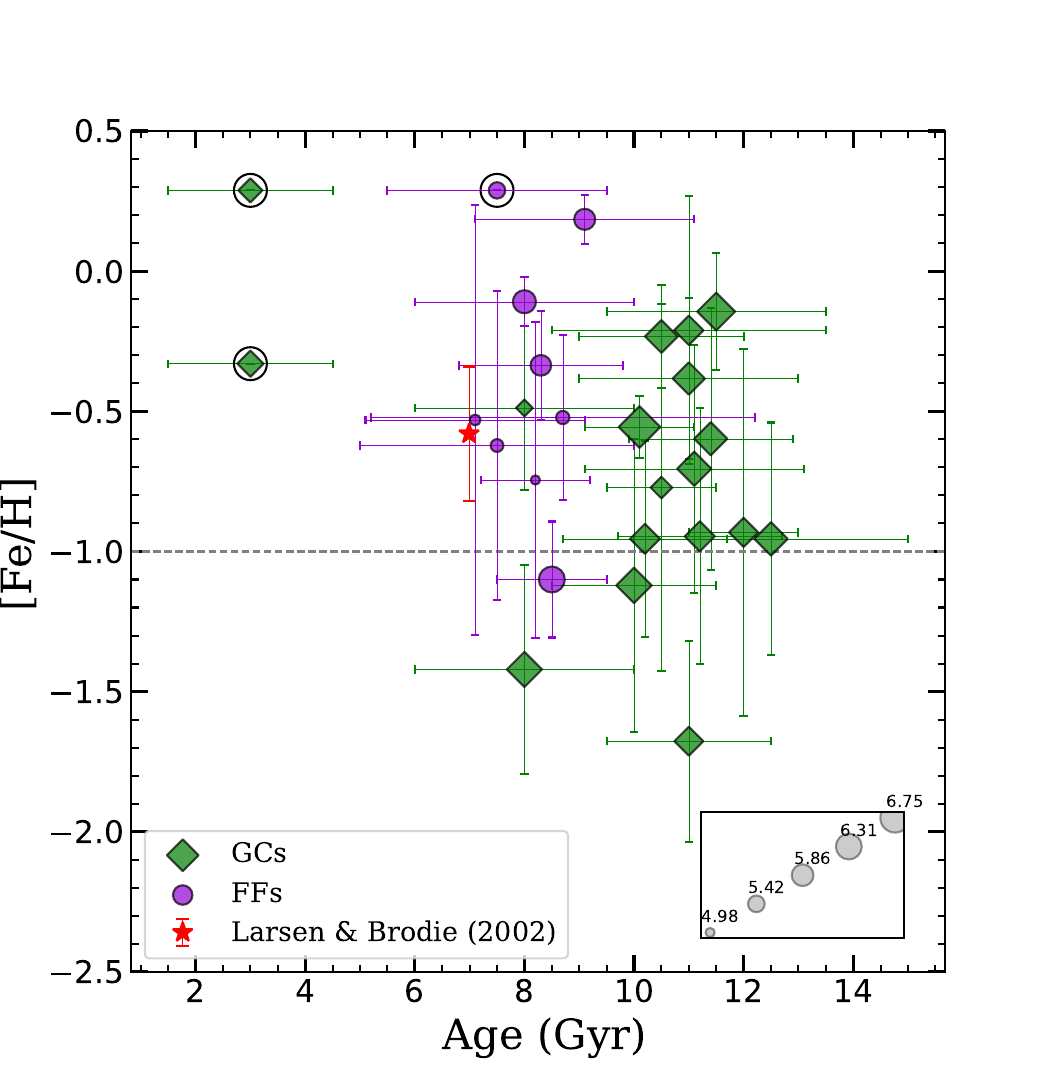}
\caption{\textit{Top:} Age distribution of GCs and FFs. \textit{Bottom:}~Age versus [Fe/H] delimiting (gray dashed line) the metal rich ([Fe/H] $>-1.0$) and metal poor ([Fe/H] $<-1.0$) GCs (green solid diamonds) and FFs (violet solid circles) in our sample.  The circled black dots in the diagram represent clusters for which metallicity was determined using the grid method. The inset bubble chart illustrates how the mass (log scale) of both GCs and FFs correlates with marker size. A specific comparison with data from \citet{2002AJ....123.1488L} is indicated by a red solid star.
\label{fig:agehisto}}
\end{figure}

\subsection{Age and metallicity of GCs and FFs in NGC\,1023}\label{sec:4.2}

The age and metallicities carry crucial information on the origin of the clusters. The measurement of the age and metallicity of GCs and FFs in NGC\,1023 is particularly significant, as this is the galaxy where FFs were identified as a distinct class of clusters for the first time. It may be recalled that the only measurements of age and metallicities of FFs come from a stacked spectra of 11 FFs from Keck observations, from which authors \citep{2002AJ....123.1488L} indicated that FFs are old, metal-poor objects much like the GCs. The availability of these quantities for a handful of individual clusters of both kinds in the same galaxy allows us to discern the different formation scenarios that had been proposed for the FFs.

The age distribution of GCs and FFs from our study is shown in the top panel of Fig.~\ref{fig:agehisto}. It can be seen that most of the classified GCs in NGC\,1023 have ages similar to the classical Galactic GCs with ages $>$10~Gyr, with the distribution showing a clear peak $\sim$11~Gyr. At the same time, we identify that four GCs are much younger than typical Galactic GCs. 
On the other hand, the ages of the FF population range between 7--9~Gyr, with the majority of them having age of 8~Gyr.

In the bottom panel of Fig.~\ref{fig:agehisto}, we show the metallicities of GCs and FFs plotted against their ages. A horizontal line at [Fe\slash H]=$-$1 is drawn to separate the clusters into metal-poor and metal-rich classes. We find that the majority of both the kinds of clusters are metal-rich. The metallicity reported by \citet{2002AJ....123.1488L} from the stacked spectrum falls in the middle of our individual measurements. On the other hand, the age inferred by \citet{2002AJ....123.1488L} is marginally lower (7 vs 8~Gyr) as compared to the mean age for our FF sample. Most of our GCs are at small galactocentric distances (see Fig.~\ref{fig:Cortesiversusours}), which could be the reason for the predominance of metal-rich bulge GCs and the relative scarcity of metal-poor halo GCs in our sample. A trend for the younger clusters to be more metal-rich could be inferred from the figure, regardless of whether the cluster is classified as FF or GC. This trend is along the expected lines in which galaxies become progressively metal-rich as more and more metals are injected into the interstellar medium by the dying stars as the galaxy ages. This suggests that FFs are bona fide star clusters formed {\it in situ} in the disk of the galaxy \citep{2010ApJ...725.2312O}.

In the bottom panel of Fig.~\ref{fig:agehisto}, the size of the symbols represents their masses, with the inset box providing a guide to the masses. Thus, FFs are systematically less massive, in addition to being systematically younger and more metallic as compared to the GCs.

\subsection{Blue Horizontal Branch stars}\label{sec:bhb}

\begin{figure}
\includegraphics[width=\columnwidth]{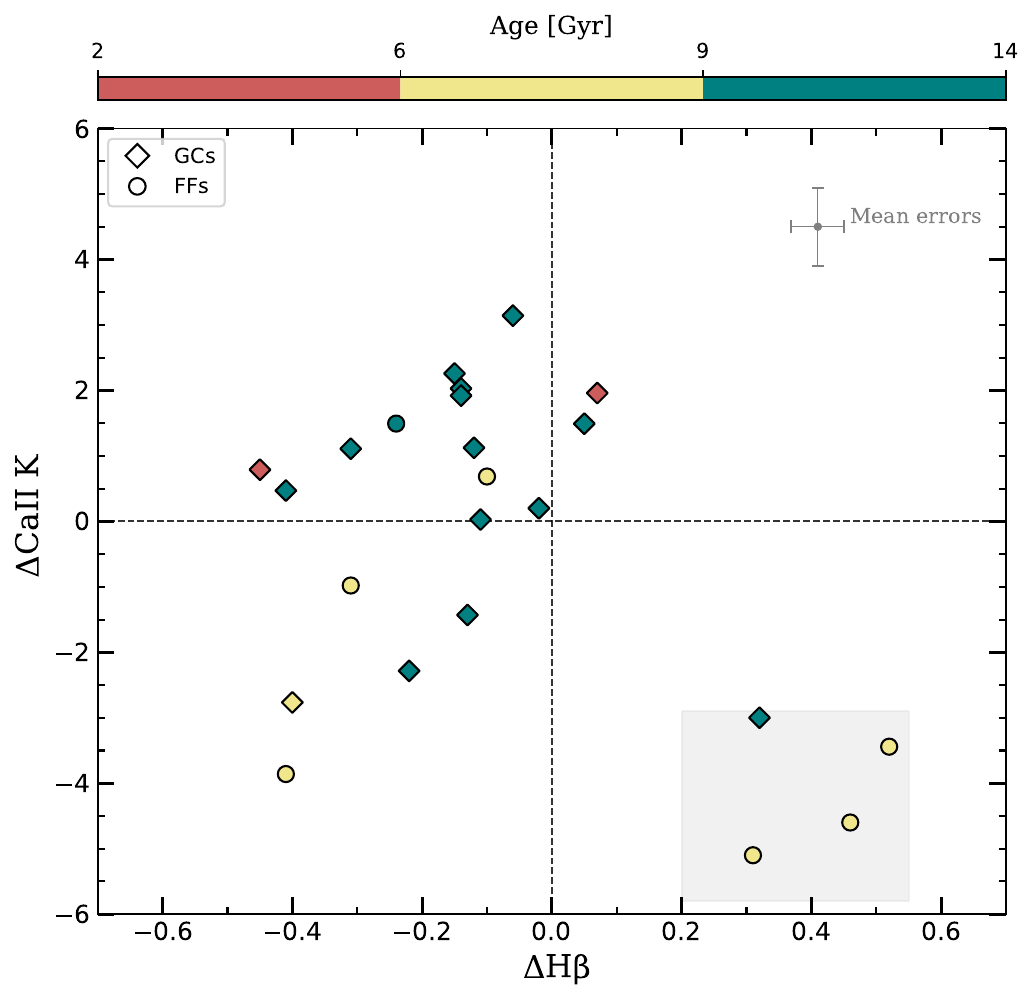}
\caption{$\Delta$H$\beta$ vs $\Delta$CaII K plot, as measured on the fit
spectra with the observed ones. GCs are presented as solid diamonds and FFs as solid circles. Each cluster is color-coded by an age range, denoted on the scale bar on top. The shadowed region on the plot, represents the clusters more likely to present a BHB, following the criterion used by \citet{2025A&A...696A..98T}, and therefore, underestimate age.} 
\label{fig:hbetavscaiik}
\end{figure}

The morphology of the horizontal branch, particularly the presence of Blue Horizontal Branch (BHB) stars, and the existence of blue stragglers can affect spectroscopic age determinations of integrated light of old stellar populations \citep{1995A&A...302..718D, 2000AJ....120..998L, 2000ApJ...541..126M}. This can lead to underestimates of ages, if the models used in the fitting process do not account for such components.
Currently, the only reliable way to distinguish between an old population containing BHB stars and a relatively  young population (age~$\leq$~9 Gyr) is through the construction of color–magnitude diagrams (CMDs). However, this method is restricted to nearby galaxies within the Local Group, where the populations can be resolved into individual stars. In unresolved stellar
populations, the presence of this hot stellar component introduces systematic errors in age determinations derived from integrated light spectra. 
Spectral features such as H$\beta$ and CaII have been widely used as diagnostics to detect the presence of a BHB morphology in unresolved stellar populations \citep{1984AJ.....89.1238R, 2004ApJ...608L..33S, 2011MNRAS.412.2445P}. 
Nevertheless, over the past years, several studies have successfully incorporated the BHB component into SSP models for MW GCs samples \citep[e.g.][]{2008MNRAS.385.1998K, 2022MNRAS.511..341C}. 
Since the SSPs used in our work do not account for BHB stars, we investigated their potential presence by following a similar approach used in \citet{2025A&A...696A..98T}, by comparing the H$\beta$ and CaII K indices measured from both the best-fitting models and the observed spectra to identify any discrepancies. The results are shown in Fig.~\ref{fig:hbetavscaiik}. We plotted the differences of the two indices as measured on the fit spectra with the observed ones. We found that most of the clusters follow a positive correlation between the differences of the two indices, however, four clusters do not follow this trend. Three FFs with an age range of 7.50-8.50 Gyr, and one GC with an age reported of 12.50 Gyr. To highlight where the differences between model and observation become significant, we have shaded the corresponding region in the figure, suggesting the presence of BHB stars in these clusters. These hot stars can significantly contribute to spectral indices, such as H$\beta$ and CaII K, in the blue part of the spectrum, and because our SSP models do not include this BHB contribution, the fitting process may compensate by returning systematically underestimated ages resulting in age determinations biased toward younger values despite the clusters being genuinely old. In Table~\ref{tab:besfittingvalues}, we have indicated which clusters may be influenced by the presence of a BHB component.

\subsection{FFs as evolved disk clusters}\label{sec:4.3}

The FFs in NGC\,1023 belong to the disk, unlike the GCs which belong to the spheroid component. Furthermore, we found that the FFs are systematically younger than the GCs. The distinct spatial location and age distribution rule out a common origin for these two classes of objects. We find that FFs, with the exception of one, are metal-rich, which excludes the possibility that they are formed during the accretion of gas-rich metal-poor galaxies as suggested by \citet{2002AJ....124.2006F}. On the other hand, their metallicity, the disk location, and the relatively younger age all suggest that they are surviving disk clusters. The relative scarcity of objects classified as GCs sharing the ages of FFs suggests that the majority of GCs that coeval to the FFs follow the same evolutionary process as them. In this section, we discuss the observed properties in this scenario of FFs as evolved disk clusters, a scenario suggested by \citet{2015ApJ...805..160S}.

It is likely that some FFs have been observed and studied in galaxies but are not recognized as a distinct category, separate from traditional GCs. The subtle distinctions between FFs and GCs, coupled with their relatively low surface brightness and diffuse nature, may have contributed to their under-reporting or misclassification. 

The disk location of FFs makes them more prone to destabilizing dynamical effects than the objects located in halos, with the principal processes of disruption being spiral arm passing and chance collision with molecular clouds \citep{2006MNRAS.371..793G}. These collisions are the main reasons for the relative scarcity of old open clusters in the Milky Way \citep{2005A&A...429..173L}. Open Clusters are low-mass and extended objects from their formation periods, which makes them susceptible to destruction. Survival of clusters for 10~Gyr in the disk requires the objects to be much more compact and massive than the Open Clusters. SSCs seen in presently active star-forming galaxies such as Antennae \citep{1995AJ....109..960W} and M82 \citep{1995ApJ...446L...1O,2008ApJ...679..404M} precisely are the objects possibly capable of surviving for long periods even under the harsh disk environment. 

Clusters experience expansion in their early phase due to expulsion of the remnant gas by the exploding supernova. Such expansion has been observed in clusters in the LMC and SMC \citep{2008IAUS..246..176M} in addition to some spiral galaxies \citep{2008MNRAS.389..223B}.
This expansion spreads the stars to larger radii, eventually leading to the escape of stars and destruction of the cluster, especially in the inner disks of giant galaxies, where the tidal effects produced from gravitational forces are stronger.
The more massive a cluster is, the higher the tidal radius under the same external forces, implying that massive clusters can hold expanding stars for longer periods of time compared to low-mass clusters.
\cite{2021MNRAS.500.4422C} carried out a study of long-time survival of SSCs in the disk of M82, a relatively low-mass disk galaxy compared to the MW. They found that the most massive and compact clusters are able to survive even up to 12~Gyr in spite of them continuously expanding and fading as more and more stars cross the tidal radius. The mass and radius of the surviving clusters are very much comparable to those of the FFs. Some expanding clusters, especially the most massive ones, could undergo core collapse at later stages in their evolution \citep{2003MNRAS.338...85M}, in which case such clusters would be classified as GCs coeval to FFs. However, the majority of the surviving clusters in the disk are not expected to be as compact and massive as the present-day classical GCs. This explains the division of FFs from GCs in magnitude vs. radius diagrams, where the FFs occupy the ``faint and large" part of the diagram.

Hence, we conclude that the FFs in NGC\,1023 and other galaxies are surviving disk clusters, whose progenitors were similar to the relatively younger SSCs found in starburst galaxies. The surviving clusters represent the massive end of the cluster mass function. The natural evolution in the disk environment is responsible for their fuzzy appearance. 
Low-mass galaxies, external parts of giant galaxies or gas-poor galaxies offer some of the ideal environments for their survival. The gas-poor nature of Lenticular galaxies, which eliminates the deadly encounters with molecular clouds, is the most likely reason for their survival in NGC\,1023. This implies that the disk of NGC\,1023 had experienced a major burst of star and cluster formation around 9~Gyr ago that formed the progenitor SSCs of the present-day FFs. This past burst of star formation is likely associated with a merger event. The presence of an interacting companion NGC\,1023A offers a testimony that the environment of NGC\,1023 even now contains candidates for a future merger event.

\section{Conclusions} \label{sec:Sec5}
We carried out a spectroscopic study of a sample of 47 red star cluster candidates, 26 of which had been classified as GC and 21 as FFs, in NGC\,1023 using the MOS mode of the OSIRIS spectrograph in the GTC. We identified 12 candidates, most of which were FF candidates, as background objects and objects with emission-line spectra. Additionally, eight other candidates were probably misclassified, as they lacked the absorption features characteristic of old stellar populations and exhibited either a rising blue continuum or had a low SNR for reliable classification.
Seventeen clusters in our sample had not previously reported radial velocity measurements, making our measurements the first for these clusters. These new measurements reinforce the conclusions from previous kinematical studies of GCs and FFs that they are associated to the spheroidal and disk components, respectively.

In this study, we present for the first time the age and metallicity measurements of individual GCs and FFs. We find distinct ages for the two populations with the majority of GCs older than 10~Gyr and FFs younger than 9~Gyr, with mean ages of 12 and 8~Gyr, respectively. We found the likely presence of a BHB component in four clusters. The omission of this hot stellar population in the models introduces a systematic effect, resulting that the derived ages for these clusters may be underestimated. Highlighting the importance of carefully considering a HB morphology especially in unresolved systems.
We determine a mean [Fe/H] = $-$0.40~$\pm$~0.29~dex for FFs, which puts them in the metal-rich category.
Our sample of GCs are also metal-rich, which is most likely due to the selection bias of not observing GCs in the outer part of the galaxy, where metal-poor halo GCs are expected to lie. The age and metallicity of the FFs that we determined are consistent with these quantities reported previously from a stacked spectrum of 11~FFs. 
All the data presented in this work are consistent with a scenario in which FFs are old surviving disk clusters. Dynamical studies of the evolution of star clusters suggest that the surviving clusters are end products of the evolution of massive and compact SSCs. Based on these, we suggest that the disk of NGC\,1023 went through a major episode of star and cluster formation around 9~Gyr ago, most likely triggered by the merger of satellite galaxy such as the one presently seen, NGC\,1023A. 

\section{Data Availability }

The fits files used in the analysis carried out in this work will be shared on reasonable request to the first author.

\section{Acknowledgments}
\begin{acknowledgments}
MALS expresses gratitude for the financial support provided by CONAHCYT through a Masters and Ph. D. research fellowship, which facilitated the completion of the work presented in this study.
We acknowledge the guidance received from Luz Itzel Álvarez Cruz and David Fernández Arenas on using pPXF during early stages of this project.
\end{acknowledgments}

\vspace{5mm}
\facilities{GTC;OSIRIS, \textit{HST}}.

\software{GTCMOS \citep{2016MNRAS.460.1555G}, pPXF \citep{2004PASP..116..138C}.
          }

\newpage
\bibliography{sample631}{}
\bibliographystyle{aasjournal}

\clearpage
\appendix
\section{Confirmed cluster candidates} 
\label{sec:bonafide}

We present the figures of the best-fit model for the bona fide clusters that met the defined selection criteria outlined in this study (see subsection \ref{sec:sepbonadis}). The clusters are sorted in numerical order, following their given slitlet number in the observation (see Table \ref{tab:tableCLASS}).  The results of the fits to the spectra to derive the ages of the bona fide clusters are discussed in detail in subsection \ref{sec:spectroscopicages} and Section \ref{sec:Sec4}. For each cluster, the observed spectra, represented in black, and the best-fit model spectra, shown in red, are plotted in each figure (see caption of Fig. \ref{fig:agebestfit} of the main text for the description of the figure). Additionally, their corresponding \textit{HST} gray scale images, taken in the $F$814$W$ filter, are included for visualization. These pictures, with a field of view measuring 5"$\times$5", capture the central region of each cluster within 1" radius for the aperture.

For a more detailed summary of the derived ages and other relevant properties of these clusters, refer to Table~\ref{tab:besfittingvalues}, which provides detailed parameters, including the [Fe/H], spectroscopic ages, photometric mass and associated uncertainties.

\begin{figure}[h!]
\includegraphics[height=0.65\textheight]{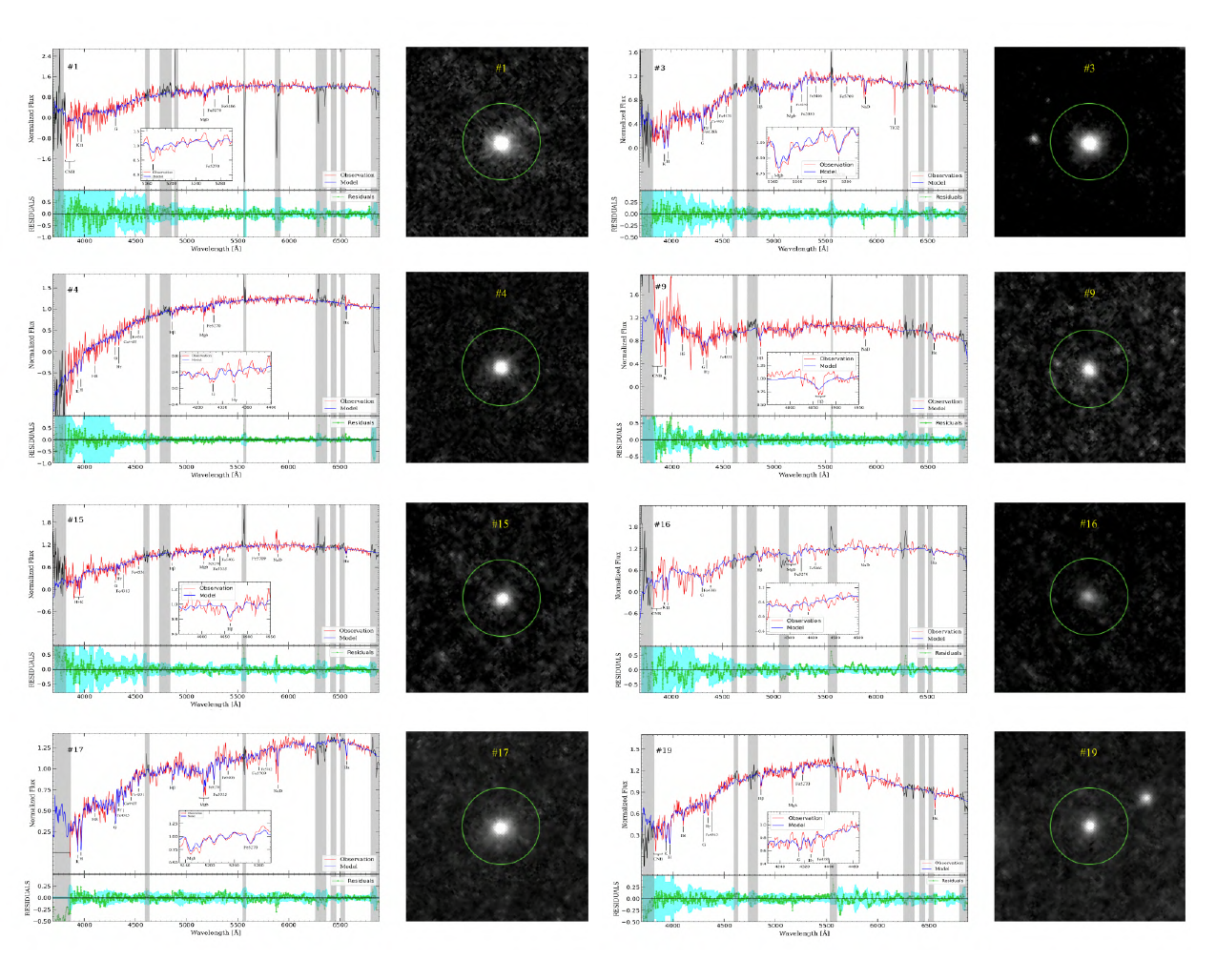}
\caption{Best-fit spectra for confirmed clusters.}
\end{figure}

\begin{figure}[ht!]
\includegraphics[height=0.96\textheight]
{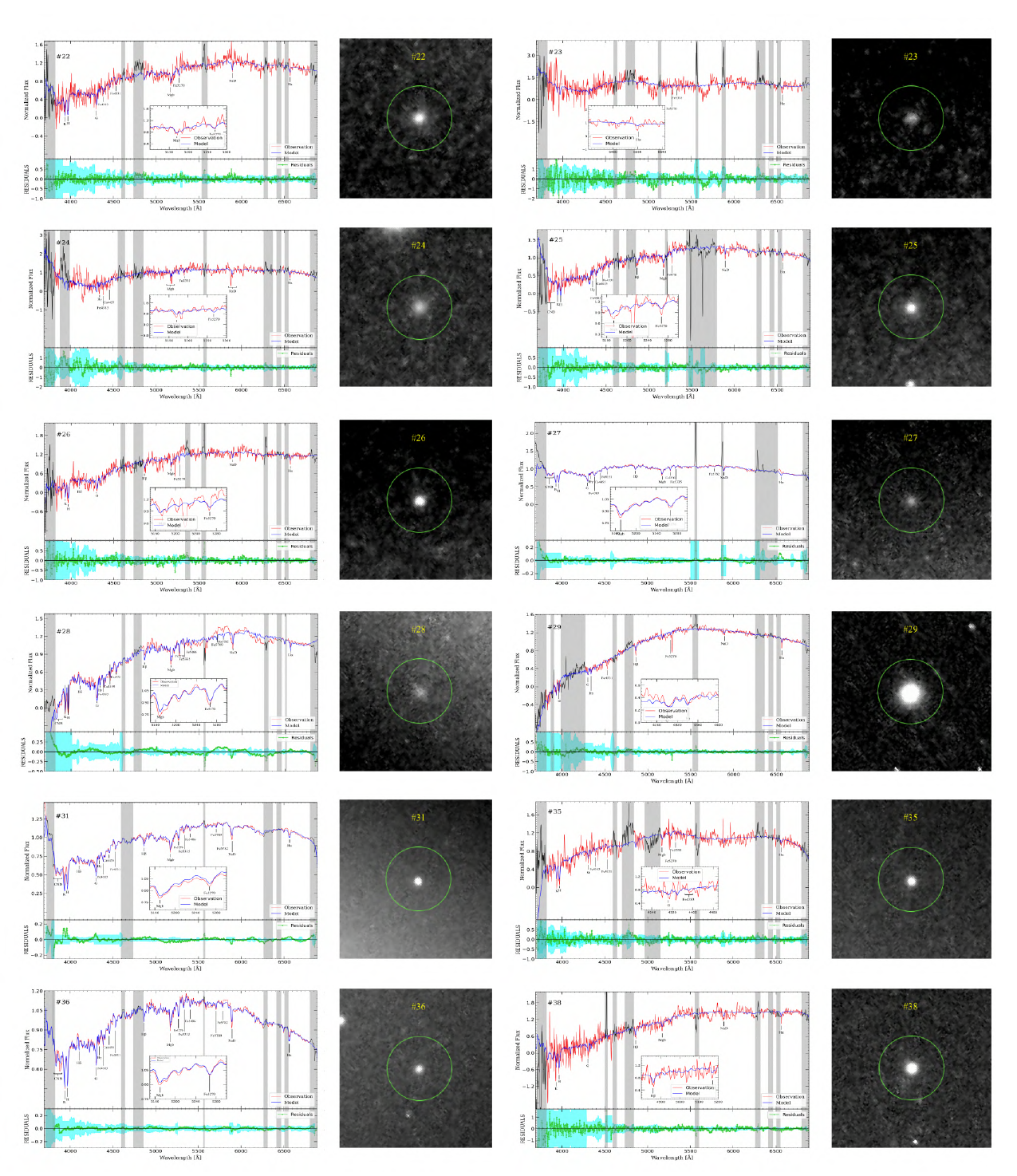}
\caption{Best-fit spectra for confirmed clusters.}
\end{figure}

\begin{figure}[ht!]
\includegraphics[height=0.5\textheight]
{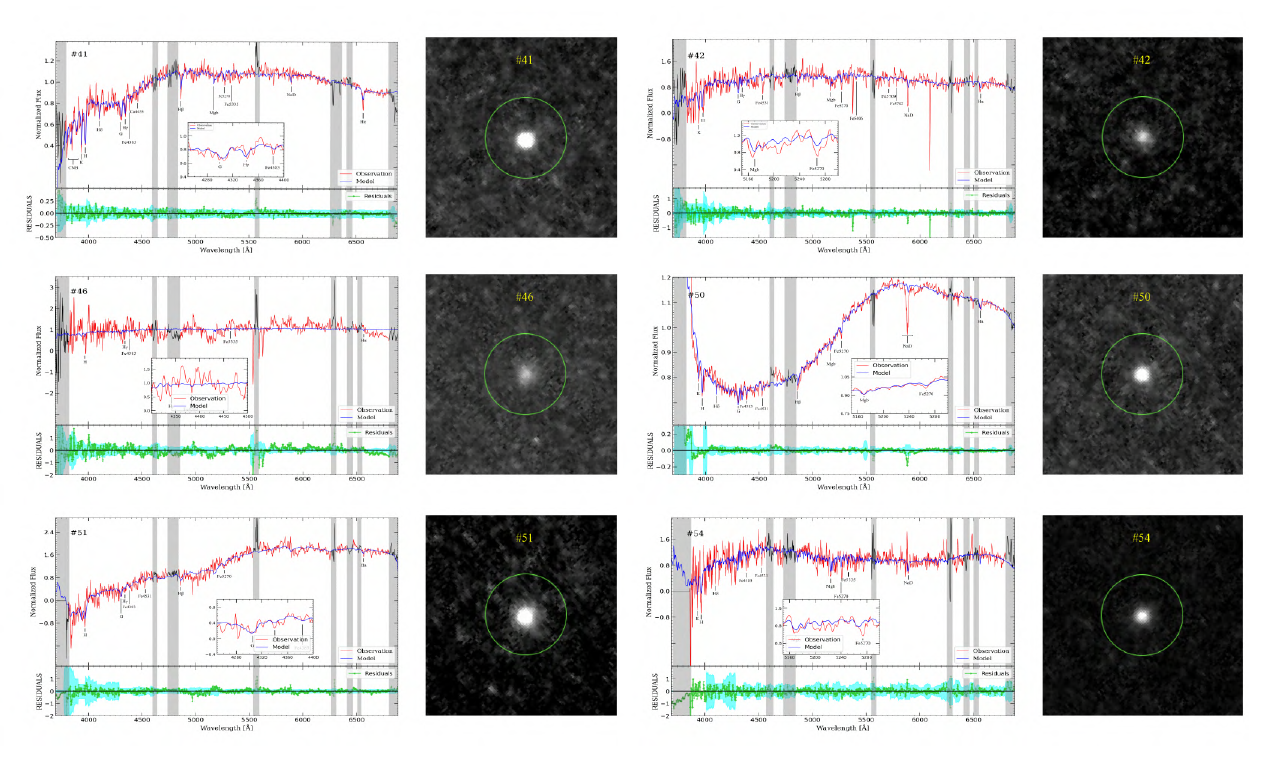}
\caption{Best-fit spectra for confirmed clusters.}
\end{figure}

\newpage
\clearpage
\section{Discarded cluster candidates} 
\label{sec:discarded}

Here, we present the discarded cluster candidates sorted in numerical order (number identified on the top left corner of each figure). These cluster candidates were excluded from the final cluster sample and classified into three main categories based on their spectral characteristics, as discussed in subsection \ref{sec:sepbonadis} and Table \ref{tab:FinalREJECTDclusters}. These sections provide additional information on the spectral and photometric characteristics that led to the exclusion of these cluster candidates from the confirmed cluster sample.

For each discarded cluster candidate, we provide its spectrum, highlighting key emission and absorption features. The skylines are marked with an "x" in each figure for clarity. Additionally, the corresponding \textit{HST} gray scale images captured using the $F$814$W$ filter are included. These images follow the same specifications and properties outlined in the previous section.

\begin{figure}[ht!]
\includegraphics[height=0.65\textheight]
{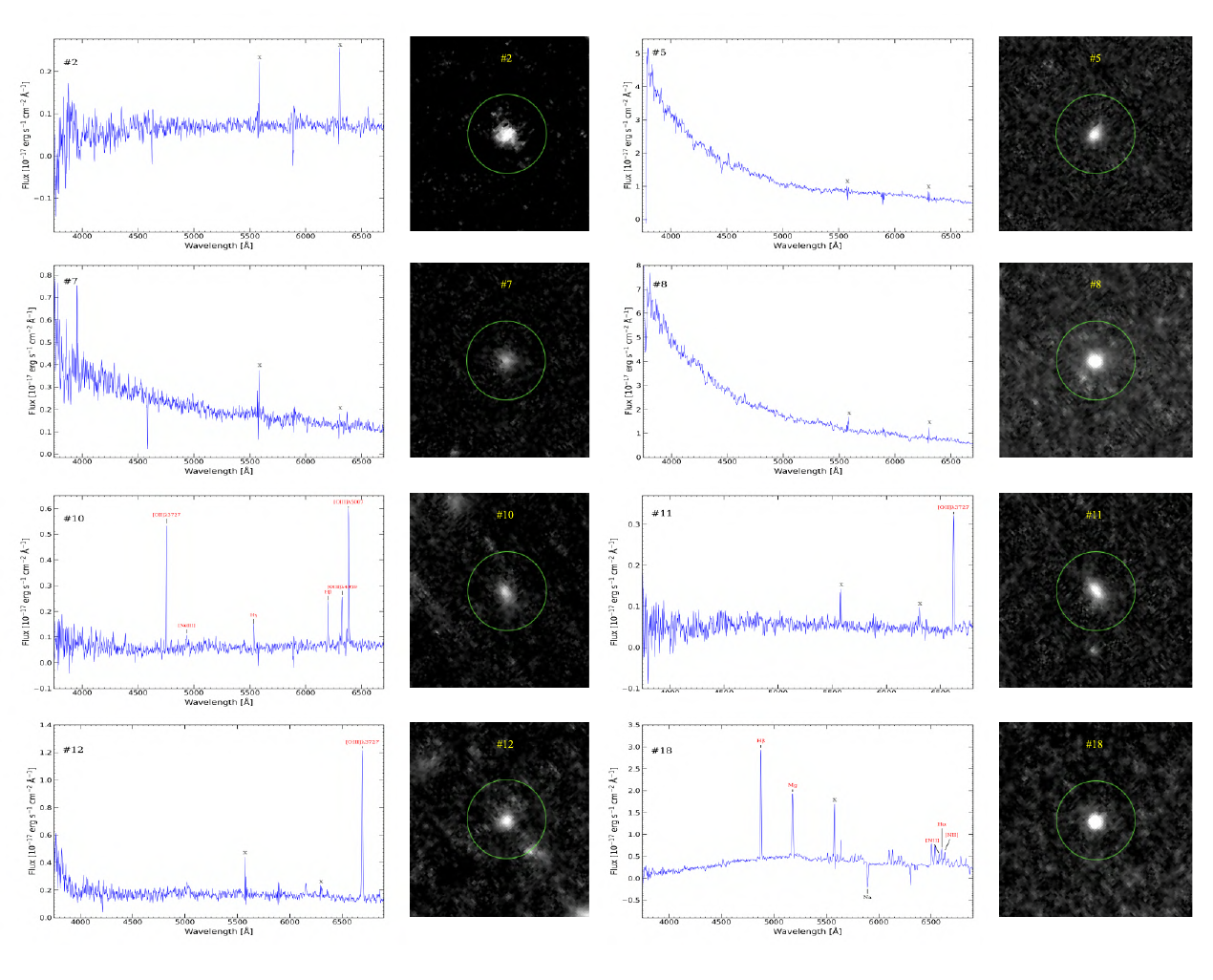}
\caption{Candidate spectra.}
\end{figure}

\begin{figure}[ht!]
\includegraphics[height=0.8\textheight]
{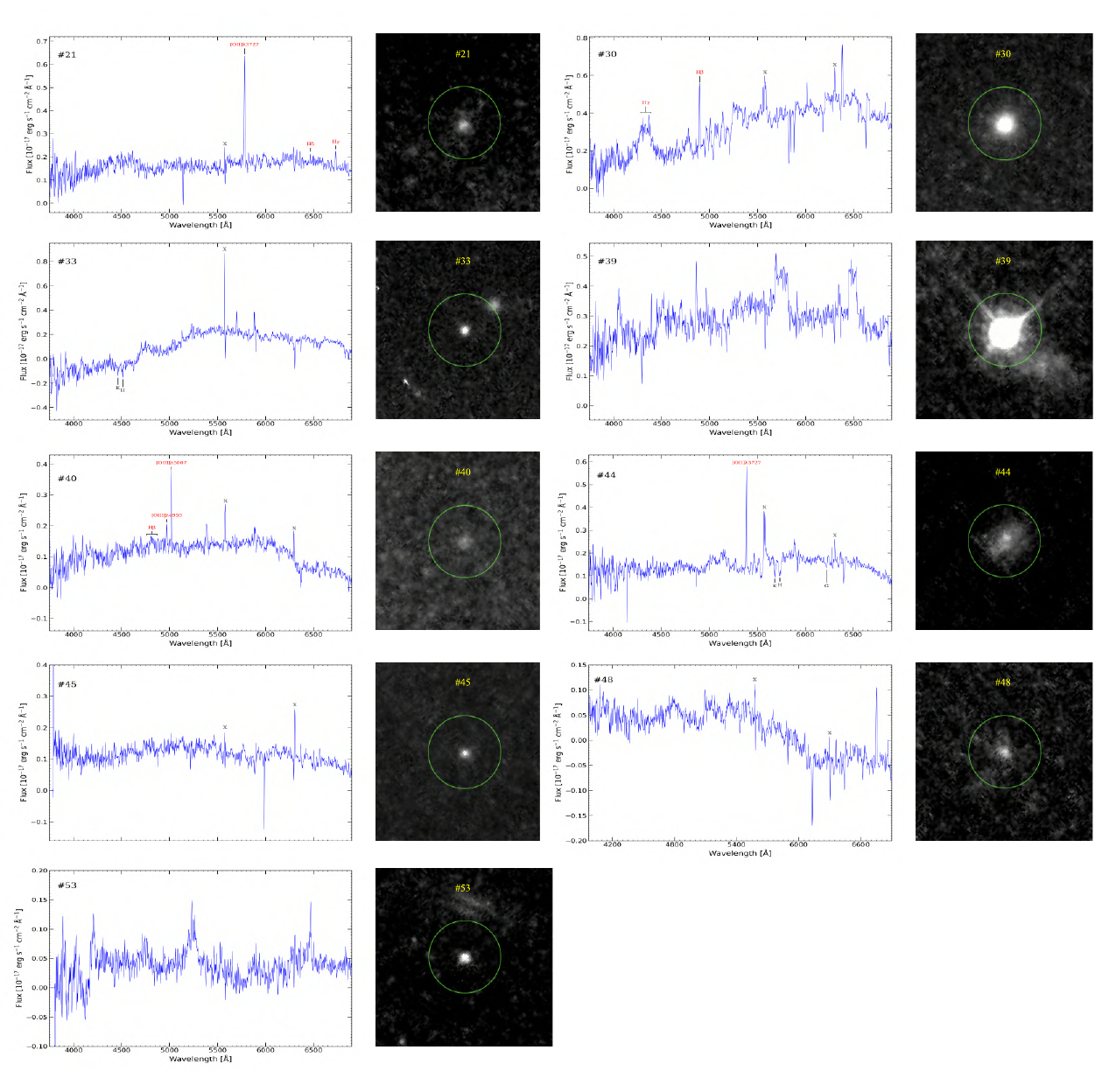}
\caption{Candidate spectra.}
\end{figure}

\end{document}